\documentclass[twocolumn,prd,preprintnumbers,nofootinbib,aps,floats,floatfix,amsmath,amssymb,secnumarabic]{revtex4} 

\usepackage{graphicx}
\usepackage[usenames,dvipsnames]{color}
\usepackage{amsmath,amssymb}
\usepackage{slashed}
\usepackage{verbatim}
\usepackage[colorlinks,citecolor=blue]{hyperref}
\usepackage[utf8]{inputenc}
\usepackage{dashbox}

\usepackage{cleveref}                    
    \crefname{equation}{}{} 
    \crefname{figure}{}{}                
    \crefname{table}{}{}
    \crefname{section}{}{}               
    \crefname{appendix}{}{}
    \crefname{footnote}{}{}
    
    \crefalias{subequation}{equation}

\usepackage[T1]{fontenc}
\usepackage{newtxtext}

\newcommand\dboxed[1]{\dbox{\ensuremath{#1}}}

\newcommand{\be}{\begin{equation}}
\newcommand{\ee}{\end{equation}}
\newcommand{\bea}{\begin{eqnarray}}
\newcommand{\eea}{\end{eqnarray}}

\def\eps{\varepsilon}
\def\eg{{\em e.g.}}

\def\phi{\varphi}

\def\lsim{\mathrel{\rlap{\lower4pt\hbox{\hskip1pt$\sim$}}
    \raise1pt\hbox{$<$}}}                
\def\gsim{\mathrel{\rlap{\lower4pt\hbox{\hskip1pt$\sim$}}
    \raise1pt\hbox{$>$}}}                

\def\Slashnew#1{#1\kern-0.55em\raise.05ex\hbox{/}}
\def\slashnew#1{#1\kern-0.5em\raise.05ex\hbox{{$\scriptstyle /$}}}

\def\S{{\scriptscriptstyle \mathrm{S} }}

\def\X{{\scriptscriptstyle \mathrm{X} }}

\def\SRS{{\scriptscriptstyle \mathrm{SRS} }}
\def\PVS{{\scriptscriptstyle \mathrm{PVS} }}
\def\GO{{\scriptscriptstyle \Gamma=0 }}
\def\eq{{\mathrm{eq} }}

\def\BW{{\scriptstyle \mathrm{BW} }}
\def\off{{\scriptstyle \mathrm{off} }}
\def\on{{\scriptstyle \mathrm{on} }}
\def\PV{{\mathrm{PV}}}
\def\l{{\scriptscriptstyle <}}
\def\g{{\scriptscriptstyle >}}

\def\sfrac#1#2{{\textstyle{\frac{#1}{#2}}}}
\def\w{{R}}

\newcommand{\Mol}{\textrm{M\o l}}

\def\Re{\mathfrak{R\!e}}				
\def\Im{\mathfrak{I\!m}}				

\begin{document}
%

\title{Anatomy of real intermediate state-subtraction scheme}

\author{Kalle Ala-Mattinen}
\email{kalle.ala-mattinen@helsinki.fi}
\affiliation{Department of Physics, University of Helsinki, 
                      P.O.Box 64, FI-00014 University of Helsinki, Finland}
\affiliation{Helsinki Institute of Physics, 
                      P.O.Box 64, FI-00014 University of Helsinki, Finland}

\author{Matti Heikinheimo}
\email{matti.heikinheimo@helsinki.fi}
\affiliation{Department of Physics, University of Helsinki, 
                      P.O.Box 64, FI-00014 University of Helsinki, Finland}
\affiliation{Helsinki Institute of Physics, 
                      P.O.Box 64, FI-00014 University of Helsinki, Finland}

\author{Kimmo Kainulainen}
\email{kimmo.kainulainen@jyu.fi}
\affiliation{Department of Physics, University of Jyv\"askyl\"a}
\affiliation{Helsinki Institute of Physics, 
                      P.O.Box 64, FI-00014 University of Helsinki, Finland}

\author{Kimmo Tuominen}
\email{kimmo.i.tuominen@helsinki.fi}
\affiliation{Department of Physics, University of Helsinki, 
                      P.O.Box 64, FI-00014 University of Helsinki, Finland}
\affiliation{Helsinki Institute of Physics, 
                      P.O.Box 64, FI-00014 University of Helsinki, Finland}

\begin{abstract}
We study the origin of the real intermediate state subtraction problem and compare its different solutions. We show that the ambiguity in subtraction schemes arises from the on-shell approximation for the 2-point functions that reduces the Schwinger-Dyson equations to the Boltzmann limit. We also suggest a new subtraction scheme which, unlike the earlier definitions, never leads to negative scattering rates. This scheme also quantifies the validity of the on-shell limit in terms of an effective one-particle weight function $\w(\Delta )$, where $\Delta$ measures the region around the resonance associated with the real state.  
\end{abstract}

\preprint{HIP-2023-14/TH}
\maketitle

%
\section{Introduction}
\label{sec:intro}
%

There is a long-standing issue in setting up consistent kinetic equations for particle distribution functions when relevant scattering processes involve real particles that also appear as intermediate states. The problem is that the resonant on-shell scattering processes overlap with the decay contributions in the kinetic equations, which results in double counting unless some subtraction procedure is established for the scattering rates. Such a procedure, referred to as the real intermediate state (RIS) subtraction, was first set up in~\cite{Kolb:1979ui,Kolb:1979qa,Fry:1980ph}. The RIS subtraction has since been discussed in various different forms~\cite{Harvey:1981yk,Luty:1992un,Cline:1993bd,Bouzas:1996jg,Plumacher:1996kc,Papavassiliou:1996zn,Pilaftsis:1997jf,Hambye:2000zs,Giudice:2003jh,Pilaftsis:2003gt,Sawyer:2003yi,Buchmuller:2004nz,Nardi:2007jp,Davidson:2008bu,Deppisch:2010fr,Kiessig:2011ga,Frossard:2012pc,Iso:2013lba,Iso:2014afa,Lavignac:2015gpa,Cline:2017qpe,Biondini:2017rpb,Dev:2017trv,Bodeker:2020ghk,Matak:2022qwc,Matak:2023zox,Ala-Mattinen:2022nuj}, and an alternative method that does not rely on the RIS subtraction was recently suggested in~\cite{Laine:2022ner}. Typically the RIS subtraction is performed at the level of Boltzmann equations, implementing some formal way to isolate and remove the on-shell part, $D_\on(s)$, from the full Breit-Wigner (BW) propagator $D_\BW(s)$, thus replacing it with the off-shell part, $D_\off(s)$, that is used for scattering amplitudes. However, this procedure is inherently ambiguous due to the ambiguity in the definition of a ``real" particle with unstable states. As a result, there is large freedom in the definition of the RIS subtraction, and different RIS subtraction schemes are even known to lead to apparently unphysical negative scattering cross sections~\cite{Cline:2017qpe}.

In this article we study and compare different solutions to the RIS problem. We also study the emergence of the problem in the context of Boltzmann equations (BE) and then from a more fundamental perspective of the Schwinger-Dyson (SD) equations for the 2-point functions. SD equations are a fully consistent setup for studying unstable particles with no notion of the real intermediate states. We will show how the problem arises when the SD equations are reduced to the spectral limit. This also makes the effect of RIS subtraction to different reaction channels evident. We then suggest a new definition for the RIS subtraction, which does not suffer from the negative cross sections. The new scheme quantifies the ambiguity in the RIS subtraction  in terms of an effective one-particle weight function $\w(\Delta )<1$ for the approximate real states, where $\Delta$ measures the size of the kinematic region around the resonance that is counted to contribute to the real particle. For the real-particle picture to be a good approximation, one should be able to choose $\Delta$, which is much smaller than the characteristic energy scale in the system, such that $\w(\Delta ) \approx 1$. Failing to satisfy these conditions signals the breakdown of the on-shell limit, and a BE network with the decay channel should not be used, or should be interpreted with due care.

The article is structured as follows: We begin by revisiting the main RIS subtraction schemes in Sec.~\cref{sec:Earlier_RIS}, where we lay out the RIS subtraction at the level of propagators via formal propagator modifications. We then apply this method to an explicit example in Sec.~\cref{sec:minimal_example} and demonstrate how the standard RIS contribution emerges from  the $s$-channel interaction. In Sec.~\cref{sec:origin-of-RIS} we study the origin of the RIS problem in the Schwinger-Dyson formalism in the case of the Yukawa theory. We show how the RIS problem arises in the on-shell limit, when the resonance overlaps with the kinematic region of interest for the problem at hand. This then motivates us to suggest our new RIS-subtraction scheme in Sec.~\cref{sec:cut_scheme}. In Sec.~\cref{sec:numerical_example} we perform numerical comparisons between different subtraction methods including our new proposal and conclude in Sec.~\cref{sec:conclusions}.

%
\section{Review of the RIS-subtraction schemes}
\label{sec:Earlier_RIS}
%

The goal of the RIS-subtraction procedure is to somehow eliminate the on-shell part from the finite-width Breit-Wigner propagator:
\begin{equation}
    D_\BW(s) = \frac{1}{s-m^2+im\Gamma}  \rightarrow D_\off (s)\,,
\end{equation}
where $s$ is the Mandelstam variable, $m$ is the mass of the propagating intermediate particle, and $\Gamma$ is its total decay width. The on-shell part $D_\on(s) = D_\BW(s)-D_\off(s)$ is then reduced to a delta function in the limit $\Gamma\rightarrow 0$. It is associated with the decay processes, while the remainder $D_\off(s)$ is thought to describe the off-shell scattering processes. For this task, different technical schemes have been proposed. Let us begin with the {\em standard RIS-subtraction} (SRS) scheme that is applied at the level of the propagator function squared.

The standard subtraction scheme is based on the simple observation that the squared BW propagator itself has a formal delta-function limit:
\begin{equation}
    |D_\BW(s)|^2 = \frac{1}{m\Gamma}\frac{m\Gamma}{(s\!-\!m^2)^2 \!+\! (m\Gamma)^2} 
    \approx \frac{\pi}{m\Gamma}\delta(s\!-\!m^2)\,,
\label{eq:|D_on|^2_known}
\end{equation}
where the last equality holds approximately for small $\Gamma$. Interpreting the right-hand side of~\cref{eq:|D_on|^2_known} as the on-shell propagator for a finite $\Gamma$ can then be used as the definition for the {\em squared} off-shell propagator:
\begin{equation}
|D^{\SRS}_\off(s)|^2 = |D_\BW(s)|^2 - \frac{\pi}{m\Gamma}\delta(s-m^2).
\label{eq:D2SSoff}
\end{equation}
This procedure works for the resonances isolated in a single channel. However, we may have for example a process with $s$- and $t$-channel contributions where the $s$-channel is resonant. Then the collision integral is proportional to
\begin{equation}
   |\mathcal{M}_\mathrm{s}|^2 + |\mathcal{M}_\mathrm{t}|^2 + 2\Re[\mathcal{M}_\mathrm{s}^*\mathcal{M}_\mathrm{t} ] \,.
\label{eq:exampleM2}
\end{equation}
The rule~\cref{eq:D2SSoff} is not sufficient to deal with the resonant propagator in the mixing term, as it only tells how to remove the on-shell part from the propagator squared contributions. 

To avoid this issue, a subtraction procedure at the level of the propagator is needed. 
This was discussed in~\cite{Harvey:1981yk} but, to our knowledge, properly proposed later in~\cite{Luty:1992un}. Here one starts by writing
\begin{align}
D_\BW(s) &= \frac{s-m^2 }{(s-m^2)^2 + (m\Gamma )^2} - i\frac{m\Gamma}{(s-m^2)^2 + (m\Gamma )^2}  
         \nonumber\\
         &= \Re\!\left[D_\BW(s)\right] + i\Im\!\left[D_\BW(s)\right]  \nonumber\\
         & \rightarrow \Re\!\left[D_\BW(s)\right] - i\pi \delta (s-m^2).
\label{eq:D_off}
\end{align}
In the last line the limit $\Gamma \rightarrow 0$ was assumed (only) in the imaginary part of the propagator. One can use this result to complete the SRS scheme for the interference term in~\cref{eq:exampleM2}, setting
\begin{equation}
D^{\SRS}_\off(s) \equiv D_\BW(s) + i \pi \delta(s-m^2)
\label{eq:DSSoff}
\end{equation}
at the propagator level, whenever a single resonant propagator (as opposed to a squared one) is encountered. In the {\em principal value subtraction} (PVS) scheme suggested in~\cite{Luty:1992un} one reverses the logic and assumes that the off-shell propagator is just the real part of the Breit-Wigner propagator:
\begin{equation}
D^{\PVS}_\off(s) \equiv \Re\!\left[D_\BW(s)\right].
\label{eq:DPVoff}
\end{equation}
Also in the PVS scheme different definitions for the propagator and for the squared propagator are needed, which require some care. One might naively think that the on-shell part of the squared propagator would be just $|D_\on|^2 \equiv |D_\BW|^2 - \left|\Re\!\left[D_\BW\right]\right|^2 \!= \left|\Im[D_\BW]\right|^2$, but this is \textit{not} the correct subtraction, because the imaginary part squared produces only half of the on-shell contribution:
\begin{align}
    \left|\Im\!\left[D_\BW\right]\right|^2  \rightarrow \frac{\pi}{2m\Gamma}\delta(s-m^2) \,. 
\label{eq:D_on_misses_half}
\end{align}
As pointed out in~\cite{Giudice:2003jh,Pilaftsis:2003gt,Sawyer:2003yi}, this issue is particularly relevant in the resonant leptogenesis literature.

The problem is that when we extract a distribution from a function the remainder is also a distribution, and one must be careful when taking the square. The explicit case at point here is that
\begin{equation}
\Big( \PV\!\Big\{ \frac{1}{x} \Big\} \Big)^2 \neq \PV\!\Big\{ \frac{1}{x^2} \Big\}\,,
\end{equation}
where PV refers to the principal value part of the function. Indeed, while $1/x$ has the principal value sequence $\PV(1/x) = \Re(1/(x+i\eps )) = x/(x^2 + \epsilon^2)$, 
the corresponding sequence for $1/x^2$ is
\begin{equation}
    \PV\!\Big\{ \frac{1}{x^2} \Big\} = \,\Re \Big( \frac{1}{(x \pm i\eps)^2} \Big)
                                        = \frac{x^2 -\eps^2}{(x^2 + \eps^2)^2}\,.
\label{eq:PVinvx2}
\end{equation}
In these expressions, the limit $\epsilon \rightarrow 0$ is of course assumed. From this exercise one infers that the correct squared propagator in the PVS scheme for a finite $\Gamma$ is 
\begin{equation}
D_\off^\PVS(s) = \frac{(s-m^2)^2 - (m\Gamma)^2}{((s-m^2)^2 + (m\Gamma)^2)^2}
\label{eq:D2PVoff}
\end{equation}
as was indeed suggested in~\cite{Cline:1993bd}. Subtracting this off-shell part from the full BW propagator one finds that
\begin{align}
    |D^\PVS_\on(s)|^2 &= |D_\BW(s)|^2 - |D^\PVS_\off(s)|^2
    \nonumber \\
    &=\frac{2(m\Gamma)^2}{((s-m^2)^2 + (m\Gamma)^2)^2} \nonumber \\
    &\rightarrow \frac{\pi}{m\Gamma}\delta(s-m^2)\,.
\label{eq:D2PVon}
\end{align}
That is, the narrow-width limit in the PVS scheme requires the use of the delta sequence $\pi\delta(x) = 2\eps^3/[x^2 + \eps^2]^2$.

The formulas for the SRS scheme~\cref{eq:D2SSoff,eq:DSSoff} and for the PVS scheme~\cref{eq:D2PVoff,eq:DPVoff} are equivalent up to order $\Gamma^2$. The PVS scheme also has a well defined $\Gamma\rightarrow 0$ limit for the off-shell propagators:
\begin{align}
  D^{\GO}_\off(s)     &\rightarrow {\rm PV} \Big\{\frac{1}{s-m^2}\Big\} \nonumber \\
  |D^{\GO}_\off(s)|^2 &\rightarrow {\rm PV} \Big\{\frac{1}{(s-m^2)^2}\Big\}.
\label{eq:Dlimitoff}
\end{align}
The limiting case~\cref{eq:Dlimitoff} can be understood as yet another subtraction scheme. We stress that the subtraction scheme dependence affects only the off-shell propagators. All schemes were designed to have the same on-shell limits:
\begin{align}
   D_\on(s)     &=  -i\pi \delta(s-m^2) \nonumber \\
  |D_\on(s)|^2  &=  \frac{\pi}{m\Gamma}\delta(s-m^2).
\label{eq:Dlimiton}
\end{align}
There is no \textit{a priori} reason to prefer one scheme over the other, although the limiting scheme~\cref{eq:Dlimitoff} is perhaps conceptually most consistent, as we shall argue later. It seems that most confusion related to RIS subtraction in the literature has resulted from a failure to realize that each scheme requires separate formulas for the off-shell propagator and the squared off-shell propagator functions. Finally, we point out that the above discussion is not restricted to $s$-channel processes. Resonances can appear also in $t$- and $u$-channels. When this happens, the subtraction should be performed as in the $s$-channel case.

A fundamental problem in all above schemes is that they can occasionally lead to {\em negative} reaction rates. This can happen because the off-shell propagators are negative near on-shell; if the scattering process is enhanced there, the integrated rates may also become negative. This is an apparently unphysical result, but it does not necessarily lead to a failure of the kinetic equation network. We shall return to these issues in Sec.~\cref{sec:numerical_example}. 

%
\begin{figure}
\includegraphics[width=0.35\textwidth]{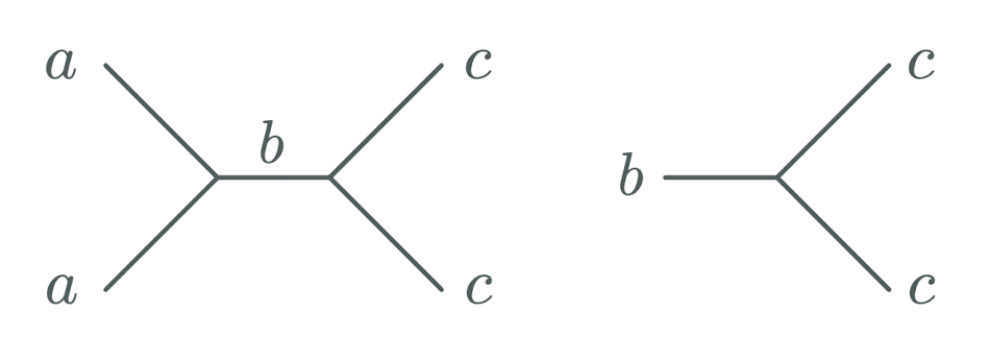}
\caption{Minimal viable set of interaction processes demonstrating a double counting of process (b) when the propagator in (a) becomes on-shell.}
\label{fig:diagrams1}
\end{figure}
%

%
\section{A minimal example of double counting}
\label{sec:minimal_example}
%
 
Here we show that the decay contribution to kinetic equations indeed corresponds to the contribution from the on-shell propagator in the scattering channel. To this end we consider a setting with unspecified particles $a$, $b$ and $c$, and for demonstrative purpose focus only on the processes shown in Fig.~\ref{fig:diagrams1}. We want to track the distribution function of species $c$, whose kinetic equation can be written as
\begin{equation}
\label{eq:BE}
    \mathrm{L}\left[f_c(p_1)\right]
    = \mathcal{C}_{aa\leftrightarrow cc}^c(p_1) + \mathcal{C}_{b\leftrightarrow cc}^c(p_1)\,,
\end{equation}
where the Liouville operator $\mathrm{L}\left[f(p)\right]  \equiv (\partial_t - H p \partial_{p})f(p)$. To allow for analytic calculation, we work in the Maxwell-Boltzmann (MB) approximation, $1\pm f_i \approx 1$ and $f_i^\eq = e^{-\beta E_i}$ for $i={a,b,c}$ and assume that particles $a$ and $b$ are in equilibrium.
We will prove that when the intermediate $b$ particle is treated at the idealized resonant limit, the equilibrium contribution of the scattering term reduces exactly to the equilibrium decay term. 

We start by evaluating the equilibrium scattering term in the MB limit:
\begin{align}
    &\mathcal{C}^{c}(a_\eq a_\eq \leftrightarrow cc)
    \nonumber \\
    &= \frac{1}{2E^{c}_{p_1}} \int_{\scriptscriptstyle p_2,k_1,k_2} \!\!\!(2\pi)^4
        \,\delta^{(4)}(k_1+k_2-p_1-p_2)\, 
    \nonumber \\
    & \;\;\;\;\times |\mathcal{M}_{aa\leftrightarrow cc}|^2\left[f_a^\eq(k_1) f_a^\eq(k_2) - f_c(p_1) f_c(p_2) \right],
\end{align}
where $\smash{\int_p \equiv \int {\rm d}^3p/[2E_p(2\pi)^3]}$. The integral over momenta $k_1$ and $k_2$ is easily performed when one demands that the detailed balance holds: $f_a^\eq(k_1) f_a^\eq(k_2) = f_c^\eq(p_1) f_c^\eq(p_2)$. Furthermore, factoring the squared matrix element as $|\mathcal{M}_{aa\leftrightarrow cc}|^2 = |\mathcal{M}_{aa\leftrightarrow b}|^2 |D_b|^2 |\mathcal{M}_{b\leftrightarrow cc}|^2$ we end up with
\begin{align}
\label{eq:C_22}
    &\mathcal{C}^{c}(a_\eq a_\eq \leftrightarrow cc)
    \nonumber \\
    &= \frac{1}{2E^{c}_{p_1}} \int_{p_2} 
      \dboxed{ \dfrac{1}{8\pi s} \lambda^{1/2}(s,m_a^2,m_a^2) 
      \,|\mathcal{M}_{aa\leftrightarrow b}|^2 |D^b_\BW|^2 }
    \nonumber \\
    &\quad \;\; \times |\mathcal{M}_{b\leftrightarrow cc}|^2\left[f_c^\eq(p_1) f_c^\eq(p_2) - f_c(p_1) f_c(p_2)  \right],
\end{align}
where $\lambda(x,y,z) = (x-y-z)^2 - 4yz$. To compare with the decay term, we need to evaluate
\begin{align}
    &\mathcal{C}^{c}(b_\eq \leftrightarrow cc)
    = \frac{1}{2E^{c}_{p_1}} \int_{p_2,q} 
        \!\!\!(2\pi)^4 \,\delta^{(4)}(q-p_1-p_2)\, 
    \nonumber \\
    &\qquad \times |\mathcal{M}_{b\leftrightarrow cc}|^2\left[f_b^\eq(q) - f_c(p_1) f_c(p_2)  \right]\,.
\end{align}
The detailed balance condition again allows us to replace $f_b^\eq(q)$ by $f_c^\eq(p_1) f_c^\eq(p_2)$. Writing the three-dimensional integral over momentum $q$ as a four-dimensional integral with the measure $\smash{\delta(q^2-m_b^2)\,(2\pi)^{-3}\mathrm{d}^4}q$  and using $\delta^{(4)}(q-p_1-p_2)$ to carry out integration over $\smash{\mathrm{d}^4\!q}$, the decay term then becomes
\begin{align}
    &\mathcal{C}^{c}(b_\eq \leftrightarrow cc)
    = \frac{1}{2E^{c}_{p_1}} \int_{p_2} 
    \dboxed{2\pi \, \delta(s-m_b^2)}
    \nonumber \\
    &\quad\;\; \times |\mathcal{M}_{b\leftrightarrow cc}|^2\left[f_c^\eq(p_1) f_c^\eq(p_2) - f_c(p_1) f_c(p_2)  \right]\,.
\label{eq:C_12}
\end{align}

Showing the equality of Eq.~\eqref{eq:C_22} and Eq.~\eqref{eq:C_12} amounts to establishing equality of the boxed quantities. To this end, we write the $|\mathcal{M}_{aa\leftrightarrow b}|^2$ matrix element in Eq.~\eqref{eq:C_22} in terms of the decay width:
\begin{equation}
    \frac{1}{8\pi} |\mathcal{M}_{aa\leftrightarrow b}|^2
    = \frac{2m_b^3}{\lambda^{1/2}(m_b^2,m_a^2,m_a^2)} \mathrm{BR}_{b\rightarrow aa} \Gamma \,,
\end{equation}
where $\Gamma \equiv \sum_i \Gamma_{b\rightarrow \{i\}}$ is the full decay width of the $b$ particle and $\smash{\mathrm{BR_{b\rightarrow \{i\}}} = \Gamma_{b\rightarrow \{i\}}/\Gamma}$ is the branching ratio to final state ${i}$. Furthermore, writing the propagator $|D^b_\BW|^2 = |D_\off|^2 + |D_\on|^2$, where the off- and on-shell results are given in~\eqref{eq:D2SSoff} and~\eqref{eq:Dlimiton}, the boxed part of the scattering term~\eqref{eq:C_22} becomes
\begin{align}
    &\dfrac{1}{8\pi s} \lambda^{1/2}(s,m_a^2,m_a^2)  
      \,|\mathcal{M}_{aa\leftrightarrow b}|^2 |D^b_\BW|^2  
    \nonumber \\
  & \rightarrow 2h(s)\,\mathrm{BR}_{b\rightarrow aa} 
       \bigg( \pi \delta(s-m_b^2) 
        + m_b\Gamma|D^b_\off|^2 \bigg) \,,
\end{align}
where we defined
\begin{equation}
    h(s) = \frac{m_b^2}{s}
           \frac{\lambda^{1/2}(s,m_a^2,m_a^2)}{\lambda^{1/2}(m_b^2,m_a^2,m_a^2)}\,.
\end{equation}
In the on-shell limit $s\rightarrow m_b^2$ the function $h(s) \rightarrow 1$ proving that
the on-shell part of the scattering term is indeed equal to the decay term contribution:
\begin{equation}
\label{eq:C22-is-BRxC12}
    \mathcal{C}^{c}(a_\eq a_\eq \leftrightarrow b_\on \leftrightarrow cc)
    = \mathrm{BR}_{b\rightarrow aa} \, \mathcal{C}^{c}(b_\eq \leftrightarrow cc)\,.
\end{equation}
This is the doubly counted state which the RIS subtraction is devised to remove via  
\begin{align}
    &\mathcal{C}^{c}(aa \leftrightarrow cc) + \mathcal{C}^{c}(b \leftrightarrow cc)
    \nonumber \\
    &\rightarrow 
    \mathcal{C}^{c}(aa \leftrightarrow b_\off \leftrightarrow cc) + \mathcal{C}^{c}(b \leftrightarrow cc) \,,
\end{align}
i.e. by dropping the on-shell part of the scattering term. Finally all 2-2 process with different initial states, including elastic channels, should be included so that their corresponding branching ratios, as in Eq.~\eqref{eq:C22-is-BRxC12}, sum up $\sum_i \mathrm{BR}_i = 1$.  

%
\section{The origin of the RIS problem}
\label{sec:origin-of-RIS}
%

The previous example verified the double counting between the decay and scattering channels, and showed that the on-shell $s$-channel propagator defined in~\cref{eq:Dlimiton} indeed corresponds to the decay contribution. In this section we shall take a look at the problem from a more general point of view.

\subsection{Spectral functions}

A simple interacting system where the RIS problem may arise is the Yukawa theory with a scalar field $\phi$ and a fermion field $\psi$. If $\phi$ is too light to decay into a fermion pair, isolated poles exist for both $\phi$ and $\psi$, and no RIS problem exists, however. In this case the vacuum scalar spectral functions $\mathcal{A}_\phi$ and $\mathcal{A}_\psi$ can be written as follows:
\begin{align}
\mathcal{A}_\phi &= \pi\epsilon(k_0) \big( R_\phi\delta(s-m_\phi^2) + \rho_\phi(s) \big)
\nonumber \\
\mathcal{A}_\psi &= \pi\epsilon(k_0)(\slashed k + m_\psi) \Big( R_\psi\delta(s-m_\psi^2) + \rho_\psi(s) \Big),
\end{align}
where $\rho_{\phi,\psi}(s)$ are the continuum (off-shell) parts to the spectral functions and the one-particle weight functions $R_{\phi,\psi}$ are constrained to be less than unity by the spectral sum rules. Indeed, $\smash{\frac{1}{\pi}\int{\rm d}k_0k_0\mathcal{A}_\phi = 1}$ and $\smash{\frac{1}{\pi}\int{\rm d}k_0\mathcal{A}_\psi = \gamma^0}$ imply that
\begin{equation}
R_{\phi,\psi} = 1 - \frac{1}{2\pi}\int {\rm d}s \rho_{\phi,\psi}(s) < 1.
\end{equation}
The continuum parts $\rho_{\phi,\psi}(s)$ may to a good accuracy be computed perturbatively using spectral free-theory propagators. In this case one can also derive a good BE approximation for kinetic equations reducing the SD equations to the on-shell limit using the spectral propagators:
\begin{align}
i\Delta_\phi^\l & = 2\pi \epsilon(k_0) R_\phi f_\phi(k) \delta(s-m_\phi^2)
\nonumber \\
iS_\psi^\l & = 2\pi \epsilon(k_0)(\slashed k + m_\psi) R_\psi f_\psi(k) \delta(s-m_\phi^2)
\label{eq:spectral_ansaz}
\end{align}
and moreover
\begin{align}
\Delta_\phi^{\rm H} & = \PV\Big\{ \frac{R_\phi }{s-m_\phi^2} \Big\}
\nonumber \\
i\Delta_\phi^{11}   & = i\Delta_\phi^{\rm H} + 2\pi \epsilon(k_0) R_\phi (f_\phi(k)+\sfrac{1}{2}) \delta(s-m_\phi^2),
\label{eq:spectral_ansaz_2}
\end{align}
where $\smash{\Delta_\phi^{11}}$ is the Feynman propagator, whose Hermitian component is given by $\smash{\Delta_\phi^{\rm H}}$. Similar equations hold for the fermion $\psi$. One can usually set $\smash{R_{\phi,\psi}=1}$ to a good approximation for perturbative couplings. We kept them here to emphasize that in an interacting theory the one-particle states do not exhaust the entire state space of the system. However, in the current setup where isolated poles exist, the on-shell formulas~\cref{eq:spectral_ansaz} and~\cref{eq:spectral_ansaz_2} provide a good parametrization for the system.

When $m_\phi > 2m_\psi$, $\phi$ is unstable and no longer has an isolated pole. In this case the spectral solutions do not strictly speaking exist, and it is not possible to derive the BE limit for the $\phi$ field without further approximations. For example, if the energy scales one is interested in do not overlap with the $\phi$ resonance, one may be able to derive a good BE limit for the problem based on the spectral parametrization for the fermion and treating $\phi$ only as an intermediate resonance. The RIS problem becomes acute only when the relevant energy scales {\em do} overlap with the resonance {\em and} one implements an on-shell kinetic equation also for $\phi$, as we shall see in the next section.

The on-shell limit can often be a good approximation also for unstable particles. When this is so, it would seem natural to use the spectral limit for all propagator functions, including the Hermitian parts. This would give rise to the limiting subtraction prescription~\cref{eq:Dlimitoff}, which in this sense is the most natural scheme to use. But this choice is not unique, as we shall see below.

%
\subsection{Schwinger-Dyson equations}
\label{sec:SD-equations}
%

We now discuss the RIS problem in the context of SD equations. We emphasize that the full SD equations, which span the entire phase space of the 2-point functions, do not need RIS subtraction. The issue emerges only when the SD equations for unstable particles are reduced to the on-shell limit. The SD equations for the Yukawa theory can be schematically written as
\begin{align}
\Delta_\phi &= \Delta_{\phi,0} + \Delta_{\phi,0} \otimes \Pi_\phi    \otimes \Delta_\phi  \nonumber \\
S      &= S_{\psi,0}      + S_{\psi,0} \otimes \Sigma_{\psi} \otimes S{\psi}
\label{eq:SD-equations}
\end{align}
where $\Delta_{\phi,0}$ and $S_{\psi,0}$ are the free propagators and $\Delta_\phi$ and $S_\psi$ are the full propagators. In the direct space representation the convolutions are defined as $(A\otimes B)(x,y) \equiv \int {\rm d}^4z A(x,z)B(z,y)$, where $z_0$ lives on the complex time contour. In real time the SD equations split into Kadanoff-Baym (KB) equations for the pole functions and for the Wightman functions. In the bosonic case we get:
\begin{equation}
\begin{split}
(\Delta_{\phi,0}^{-1} - \Pi_\phi^p) \otimes \Delta_\phi^p &= 1 \\
(\Delta_{\phi,0}^{-1} - \Pi_{\phi}^{\rm H}) \otimes \Delta_\phi^s 
               &= \Pi_\phi^s \otimes \Delta_\phi^{\rm H} + \mathcal{C}_\phi^s,
\label{eq:KB_fermions}
\end{split}
\end{equation}
where $p=r,a$ refer to the retarded and advanced pole functions and $s= <,>$ to the statistical Wightman functions, and we defined the collision terms
\begin{equation}
\mathcal{C}_\phi^\l = \frac{1}{2}\big( \Pi_\phi^\g \otimes \Delta_\phi^\l 
                                      -\Pi_\phi^\l \otimes \Delta_\phi^\g \big),
\label{eq:SDC}
\end{equation}
and $\smash{\mathcal{C}_\phi^\g = -\mathcal{C}_\phi^\l}$. Similar decomposition can be derived for the fermionic correlation functions $S_\psi^{p,s}$; see \eg~\cite{Juk21}. The KB equations usually cannot be solved without further approximations, such as the spectral ansatz~\cref{eq:spectral_ansaz}. The reduction of the KB equations to the spectral limit is a delicate task~\cite{Juk21,Kainulainen:2023ocv,Kainulainen:2023khg}. However, to understand the RIS problem, it is sufficient to concentrate on 
the collision terms ${\cal C}_{\phi,\psi}^\l$, which in the end emerge as the collision integrals in Boltzmann equations.

%
\begin{figure}[t]
\includegraphics[width=0.38\textwidth]{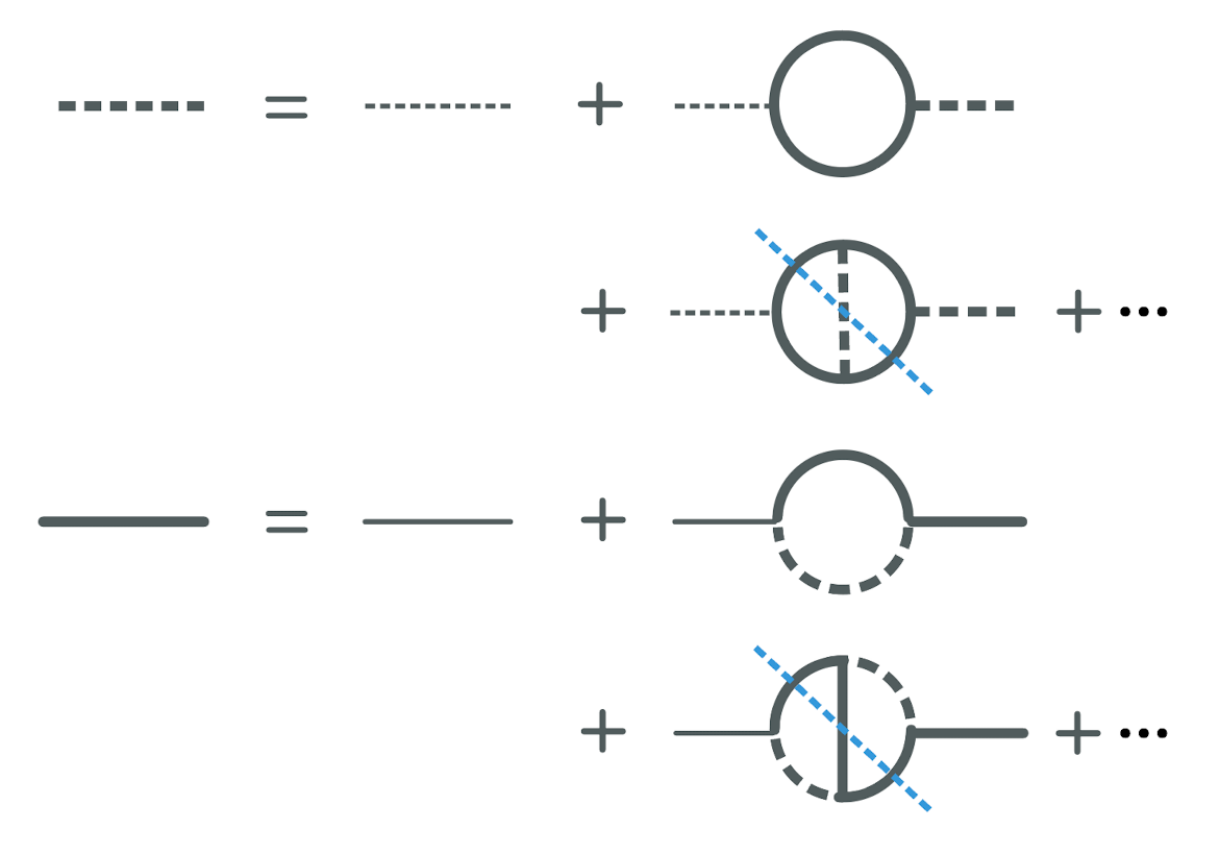}
\vskip-0.2cm
  \caption{The diagrams contributing to the SD equations in the Yukawa theory to the two-loop order in the 2PI expansion. Thin lines indicate the free and thick lines the full propagators. Blue dashed lines express cuts defined similarly to Fig.~\cref{fig:gamma2-expansion2}.}
  \label{fig:gamma2-expansion1}
\end{figure}
%

\begin{figure}[t]
\includegraphics[width=0.33\textwidth]{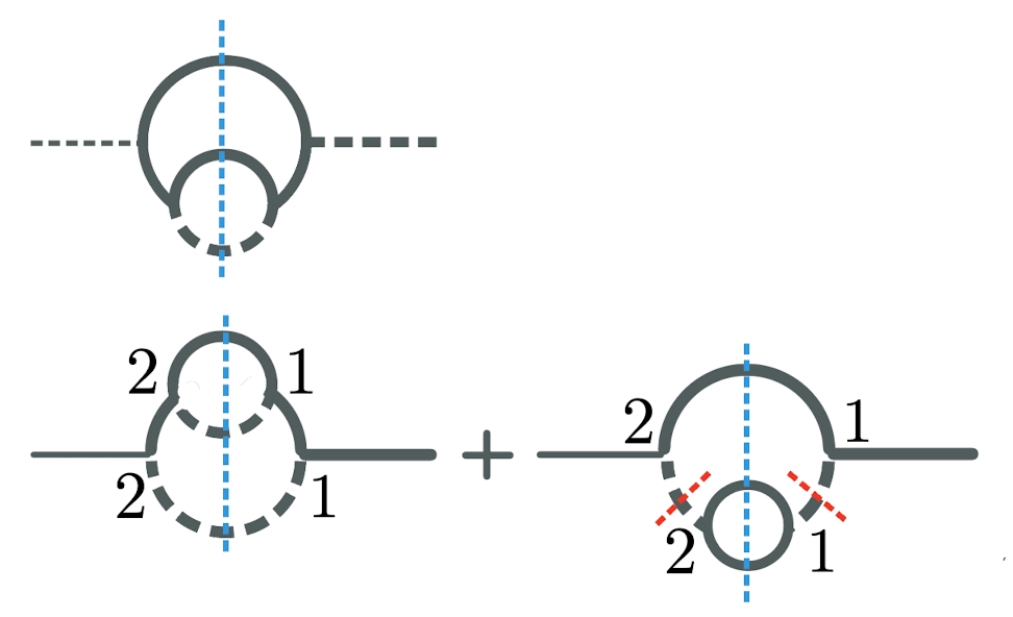}
\vskip-0.3cm
  \caption{Additional perturbative diagrams needed in the on-shell limit. The numbers are the closed time path indices singling out the embedded self-energy diagram $\Sigma^> = \Sigma^{21}$, and blue dashed lines indicate cuts corresponding to these indices. Red dotted lines in the internal boson propagators $\Delta_{11}$ and $\Delta_{22}$ indicate that these propagators are included without their on-shell parts.}
\label{fig:gamma2-expansion2}
\end{figure}

In Fig.~\cref{fig:gamma2-expansion1} we show the diagrams contributing to  Eq.~\cref{eq:SD-equations} up to two loops in the two-particle irreducible (2PI) expansion. Dashed lines correspond to $\phi$-propagators and continuous ones to $\psi$-propagators, and thin (thick) lines correspond to the free (full) propagators. We emphasize that the two-loop diagrams shown in Fig.~\cref{fig:gamma2-expansion2} are {\em not} included in full SD equations. They are already accounted for by the one-loop diagrams, because the full SD equations sum all perturbative corrections associated with the diagrams included in the 2PI expansion. However, when simplifying approximations are imposed, the consistency of the 2PI expansion breaks down and some non-2PI diagrams may need to be inserted by hand.
The simplest example is the case where we treat the $\phi$-field as a resonance and drop it from the SD equation network. In this case the scalar decay diagrams are not summed by the SD equations and the last diagram in Fig.~\cref{fig:gamma2-expansion2} must be included perturbatively into the SD equation for the $\psi$-field. 

The spectral limit is more delicate. Here the SD equation for the $\phi$-field is kept, but all collision integrals are reduced to the on-shell limit, where the one-loop terms reduce to the decay terms in the BEs. While the BEs still sum these on-shell contributions to all orders, this summation completely misses all off-shell contributions. The 2PI expansion is again broken and {\em the off-shell parts} of the diagrams shown in Fig.~\cref{fig:gamma2-expansion2} must be included perturbatively. The RIS problem has thus entered.

The cuts in two-loop diagrams indicated by blue dashed lines in Fig.~\cref{fig:gamma2-expansion1} give only the interference terms (\eg~between the $s$- and $t$-channels) in the collision integrals, while the squared matrix elements in each channel emerge from the cuts in the forbidden diagrams of Fig.~\cref{fig:gamma2-expansion2}. The summation argument applies to all noncut internal propagators, and so the need for subtraction concerns all channels and all matrix elements including the interference terms in collision integrals. In the present example the on-shell condition is kinematically forbidden from appearing in fermion lines, but it affects the scalar propagators in the last diagrams in Figs.~\cref{fig:gamma2-expansion2,fig:gamma2-expansion1}.

%
%

The standard on-shell reduction of the SD equations to BEs explicitly uses the spectral propagators of Eqs.~\cref{eq:spectral_ansaz,eq:spectral_ansaz_2}. However, one {\em can} replace the Hermitian principal value propagators by other off-shell propagators with no change to the reduction procedure. Indeed, the different subtraction schemes discussed so far, are but different definitions for the Hermitian propagator functions in the spectral ansatz. From this point of view all schemes are equally good. In the next section we will introduce another subtraction scheme that is free of the issue of negative rates and provides a quantitative estimate for the validity of the BE limit.

%
\section{Cut-subtraction scheme}
\label{sec:cut_scheme}
%

In the previous section we saw that both the need for and the ambiguity of the RIS subtraction arise when the full solutions to the SD equations are approximated by spectral solutions. The validity of the spectral limit clearly depends on the resonance width $\Gamma$. For a small width, excitations around the resonance may give nearly identical contributions to physical processes, allowing them to be treated as an effective single-particle state. If $\Gamma$ is very small, the integrated weight of such states may even saturate the spectral sum rule. This is the case when the usual spectral reduction of the SD equations to the BE limit with free theory normalization $R=1$ is valid. If the physical process changes significantly over the resonance region, however, the spectral approximation starts to break down. 

We quantify these simple observations by defining the following cut-subtraction scheme:
\begin{equation}
D^{\Delta}_\off(a) = \big[1-\Theta(a\!-\! m^2,\Delta)\big] D_\BW(a),
\label{eq:cut_scheme}
\end{equation}
where $a = s,t$, or $u$, depending on the channel one is looking at, and the cut function $\Theta$ singles out a region around the resonance. The simplest choice is the top-hat function:
\begin{equation}
\Theta(x,\Delta) =  \theta(\Delta-x) \theta(x+\Delta).
\label{eq:cut-function}
\end{equation}
The squared off-shell propagator does not need a separate rule in this scheme. The propagator in Eq.~\cref{eq:cut_scheme} is again assumed to replace the Hermitian parts in the Feynman and anti-Feynman propagators in spectral decomposition, \eg~in Eq.~\cref{eq:spectral_ansaz_2}. Combined with this prescription, the statistical propagators in Eq.~\cref{eq:spectral_ansaz} (as well as  the spectral parts of the Feynman and anti-Feynman propagators), are rescaled by a weight function: 
\begin{equation}
    \w(\Delta)
    \equiv
    \frac{1}{\pi} \int \mathrm{d}s  \, 
    \Theta(s\! -\! m^2,\Delta) \, m \Gamma |D_\BW(s)|^2.
\label{eq:weight_function_def}
\end{equation}
It should be obvious that none of these changes interfere with the on-shell reduction of the SD equations.

The precise form of the cut function in Eq.~\cref{eq:cut-function} is not relevant, and the characteristic width $\Delta$ will depend on the problem. The weight function will give a {\em quantitative measure} for the validity of the spectral limit. To see how this works, consider some physical quantity $\mathcal{F}$, which is an integral over, say, the Mandelstam variable $s$:
\begin{equation}
\mathcal{F} = \int \mathrm{d}s \, F(s) \,|D_\BW(s)|^2  \,.
\label{eq:full_F_int}
\end{equation}
For example the vacuum contribution to the $\psi$ collision integral from process $\psi\bar \psi\rightarrow \psi\bar\psi$, coming from the cut in the last diagram in Fig.~\cref{fig:gamma2-expansion2} would have this form. We can now divide $\mathcal{F}$ into on- and off-shell contributions $\mathcal{F} = \mathcal{F}_\on + \mathcal{F}_\off$, where
\begin{equation}
    \mathcal{F}_\off = \int \mathrm{d}s \, F(s) \,|D^\Delta_\off(s)|^2  \,.
\label{eq:off-shell_cut}
\end{equation}
and
\begin{align}
   \mathcal{F}_\on
    &= \frac{1}{m\Gamma} \int \mathrm{d}s \,
    F(s) \Theta(s\!-\!m^2,\Delta) m\Gamma |D_\S(s)|^2 \nonumber \\
    &\approx \w(\Delta) \pi \frac{F(m_\S^2)}{m_\S\Gamma_\S}\,.
\label{eq:on-shell_cut}
\end{align}
In the second line we assumed that $F(s)$ remains essentially a constant in the cut region and then used Eq.~\cref{eq:weight_function_def}. This clearly corresponds to replacing $|D_\BW(s)|^2$ inside the on-shell region with a weighted delta function: $\pi\w(\Delta)\delta(s-m^2)/(m\Gamma)$.

If $\mathcal{F}$ indeed described the process $\psi\bar \psi\rightarrow\psi\bar\psi$, then the on-shell part in Eq.~\cref{eq:on-shell_cut} would be exactly canceled by the on-shell vacuum decay term $\phi \rightarrow \psi \bar\psi$ arising from the one-loop diagram in the fermion SD equation in Fig.~\cref{fig:gamma2-expansion1}. Indeed, because of our rule for weighting the statistical propagators by $\w(\Delta)$, the latter is given by the standard vacuum decay term multiplied by $\w(\Delta)$. The exact calculation equating the two is essentially the same we presented in Sec.~\cref{sec:minimal_example}, with $b = \phi$ and $aa,cc\rightarrow \psi\bar\psi$.  

Our scheme in corresponds to setting $\smash{\Delta^{\rm H}_\varphi \rightarrow D^\Delta_\off}$ {\em and} $R_\phi \rightarrow \w(\Delta)$ in Eqs.~\cref{eq:spectral_ansaz} and~\cref{eq:spectral_ansaz_2}. All collision integrals can then be computed from these functions following standard finite temperature field theory methods. One then recovers the standard tree level rules for computing collision integrals, augmented with the scaling of the phase space factors $f_\phi \rightarrow \w(\Delta) f_\phi$ and replacing $D_\BW \rightarrow D^\Delta_\off$ for all intermediate scalar propagators. Extensions to more complicated theories where also fermions can be unstable should be obvious.

We should emphasize one important difference between our scheme and the other subtraction schemes. Consider the case where $F$ is a constant. In this case all standard schemes give $\mathcal{F}_\off = 0$, whereas in the cut scheme $\mathcal{F}_\off = F(1-\w(\Delta))$. The latter one is the qualitatively correct result. The standard schemes thus {\em overestimate} the weight of the effective one-particle states, which forces the flat-weight integrals of the off-shell parts to vanish in compensation. The effect underestimates the off-shell contributions also for a nonconstant $F(s)$ and may even lead to a negative cross section if the corresponding $F(s)$ is enhanced near the resonance. The cut-subtraction scheme does not suffer from these problems by construction.

The particle picture makes sense only if the approximation in Eq.~\cref{eq:on-shell_cut} holds with a large enough $\Delta$, such as $\w(\Delta)\simeq 1$. Indeed, if $F(s)$ changed rapidly in the scale $m\Gamma$, we would be forced to use very small $\Delta \ll m\Gamma$ to extract $F(s)$ outside the integral. This would then give $\w(\Delta) \ll 1$, showing that the one-particle contribution to the process is very small. Of course the particle picture would not make sense in this limit. Nevertheless, the cut-subtraction procedure allows for a continuous mapping between the limit where the particle picture is valid [$\w(\Delta) \simeq 1$] and the resonance limit where $\phi$ no longer is part of the thermal bath [$\w(\Delta) \simeq 0$].

%
\section{Numerical examples}
\label{sec:numerical_example}
%

In this section we compare quantitatively the predictions of the SRS and PVS schemes and study the cut scheme as a function of $\Delta$ in a physical system where RIS subtraction is needed, at least in principle. To be specific, and to keep the discussion as simple as possible, we consider the dark matter freeze-out in the singlet extension of the standard model(~\eg~\cite{Silveira:1985rk,McDonald:1993ex,Burgess:2000yq,Barger:2007im,Cline:2013gha}), where the DM-abundance calculations have been recently studied to high precision~\cite{Binder:2017rgn,Ala-Mattinen:2019mpa,Ala-Mattinen:2022nuj,Laine:2022ner,Kim:2023bxy}. The model is described by the Lagrangian
\begin{equation}
  \mathcal{L} =
   \sfrac{1}{2}(\partial_\mu S)^2 - \sfrac{1}{2}\lambda_{hs} h^2S^2 + \dots \,,
\label{eq:symm_lagrangian}
\end{equation}
where $h$ is the Standard Model Higgs, $S$ is a singlet scalar, and dots refer to other terms whose precise form is not relevant here, including the SM Lagrangian. We focus on the $h$ and $S$ particle densities governed by the following set of coupled Boltzmann equations:
\begin{align}
\begin{split}
    \mathrm{L}\left[f_h(p_1)\right]
    &= 
    \mathcal{C}_{h\leftrightarrow SS}^h(p_1) + \mathcal{C}_{h\leftrightarrow \rm (SM)}^h(p_1) +
    \mathcal{C}_{\rm other}^h \,,
    \\
    \mathrm{L}\left[f_S(k_1)\right]
    &= 
    \mathcal{C}_{SS\leftrightarrow \rm h}^S(k_1) + \mathcal{C}_{SS\leftrightarrow \rm (SM)}^S(k_1) + \mathcal{C}_\mathrm{other}^S \,.
\label{eq:coupled_BE}
\end{split}
\end{align}

The scattering process $SS\leftrightarrow \rm (SM)$ describes the annihilation of the $S$ scalars to the standard model (SM) particles via $s$-channel Higgs boson resonance. Other channels that could affect these densities, not relevant for the present discussion, are contained in $\mathcal{C}^{h,S}_\mathrm{other}$. The RIS subtraction is necessary in this model, because the decay and inverse decay terms $h \leftrightarrow SS$ and $h \leftrightarrow \rm (SM)$ already account for the on-shell part of the scattering term as discussed in previous sections.

To perform a full comparison we should solve Eq.~\cref{eq:coupled_BE} using the different schemes for the off-shell propagator in the scattering integral. In the momentum dependent case the collision integrals should be arranged to contain an integral over $s$, as was done \eg~in~\cite{Ala-Mattinen:2019mpa}, to facilitate the cut regularization. We will instead concentrate on the simpler momentum integrated version of Eq.~\cref{eq:coupled_BE}, where the relevant dynamics are given by the thermal averaged cross section~\cite{Gondolo:1990dk}\footnote{
In thermal equilibrium this rate is exact when $S$ scalars are treated in spectral approximation and we assume the MB statistics. Indeed, in thermal equilibrium the full solution to the Higgs SDE is $i\Delta_h^\l = 2f_\eq{\cal A}_h$, where ${\cal A}_h = \epsilon(k_0)m\Gamma D_\BW(s)$. With these assumptions the one-loop collision term in the SD equation for $S$ is given by Eq.~\cref{eq:C_12}, with the delta function replaced by $(m\Gamma/\pi)D_\BW(s)$. Tracing the derivation in Sec.~\cref{sec:minimal_example} backwards, one can establish the equality of that result and~\cref{eq:C_22} which, when integrated over the initial momentum, gives~\cref{eq:vsg} with the BW propagator. From this result one then must subtract the on-shell part, which eventually gives~\cref{eq:vsg}.
} 
\begin{align}
    \langle v_\Mol \sigma_\X \rangle 
    &=
    \frac{1}{8 m_\S^4 TK^2_2(\frac{m_\S}{T})}
    \int_{4m_\S^2}^\infty \nolimits \! \mathrm{d}s
    s^{3/2}v^2_\S(s)
    K_1\!\Big(\frac{\sqrt{s}}{T}\Big) \sigma_\X (s)\
    \nonumber \\
    &\equiv 
    \int_{4m_\S^2}^\infty \nolimits \mathrm{d}s  F(s) |D^\X_h(s)|^2.
\label{eq:vsg}
\end{align}
Here $v^2_\S = 1-4m_\S^2/s$ and $K_i(x)$ are the modified Bessel functions of the second kind. In the second line we extracted the subtracted propagator function, letting $F(s)$ describe the rest of the integrand, and the index X refers to the subtraction scheme.

%
\begin{figure}
\includegraphics[width=0.38\textwidth]{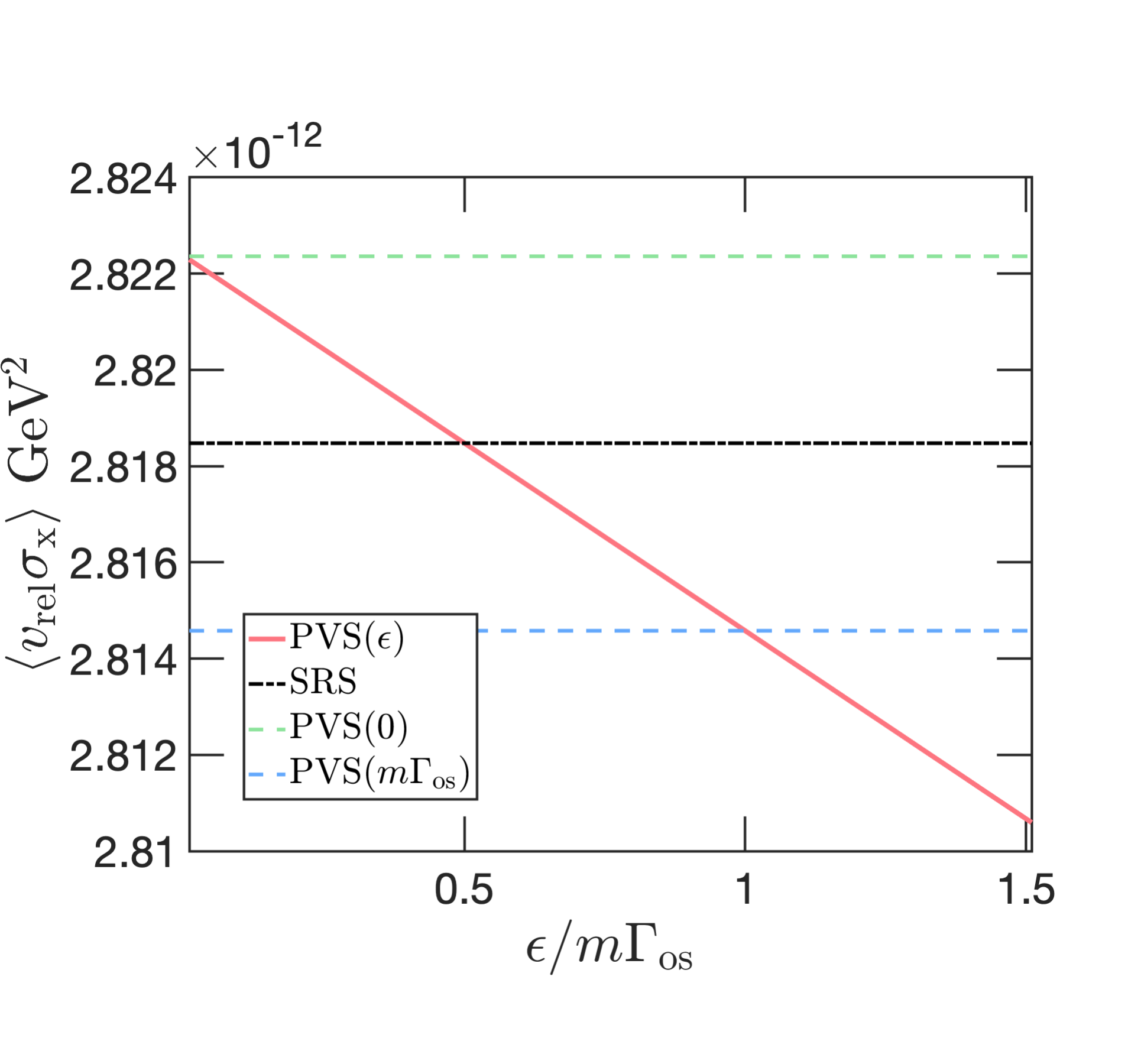}
\vskip-0.3cm
\caption{Comparison of the off-shell parts of the thermal averaged cross section in Eq.~\cref{eq:vsg} for the standard subtraction schemes. We used $m_h = 125$ GeV, $m_S = 50$ GeV, $\lambda_{hs} = 0.001$, and $T = 5$ GeV.}
\label{fig:offshell_comparison}
\end{figure}
%

The $s$-channel cross section, from which $F(s)$ can be read off using Eq.~\cref{eq:vsg}, is given by~\cite{Cline:2013gha}
\begin{equation}
\sigma_\X(s) = \frac{y^2(\lambda_{hs}v)^2}{v_\S \sqrt{s}} | D^\X_{\rm S}(s)|^2 \Gamma_h(\sqrt{s}),
\end{equation}
where $v=246$ GeV is the SM-Higgs vacuum expectation value and $\Gamma_h= \Gamma_{\S\S}+\Gamma_{\rm SM}$ is the total width of a virtual Higgs particle with an effective mass $\sqrt{s}$. The rate for $h\rightarrow SS$ is $\Gamma_{\S\S} = (\lambda_{hs}v)^2v_\S/(32\pi\sqrt{s})$. For the exact definition of $\Gamma_{\rm SM}$ see~\cite{Cline:2013gha}.

Figure~\ref{fig:offshell_comparison} shows the $s$-channel thermal average $\langle v_\Mol \sigma_\X \rangle$ in different subtraction schemes for a particular set of parameters given in the figure caption. The solid red line shows the PVS-subtracted propagator as a function of a varying effective width $\epsilon$. The dash-dotted black line is the SRS-subtracted result corresponding to Eq.~\cref{eq:D2SSoff}, while the blue dashed line shows the result for the PVS propagator~\cref{eq:D2PVoff} with $\epsilon = \epsilon_{\rm os} \equiv m_h\Gamma_{\rm os}$, and the green dashed line shows the spectral PV propagator, Eq.~\cref{eq:Dlimiton}. We take the area between the dashed lines corresponding to two constant width PVS schemes as indicative of the fundamental ambiguity in the calculation of the scattering rate using the SRS- and PVS-subtraction schemes.

%
\begin{figure}
\includegraphics[width=0.38\textwidth]{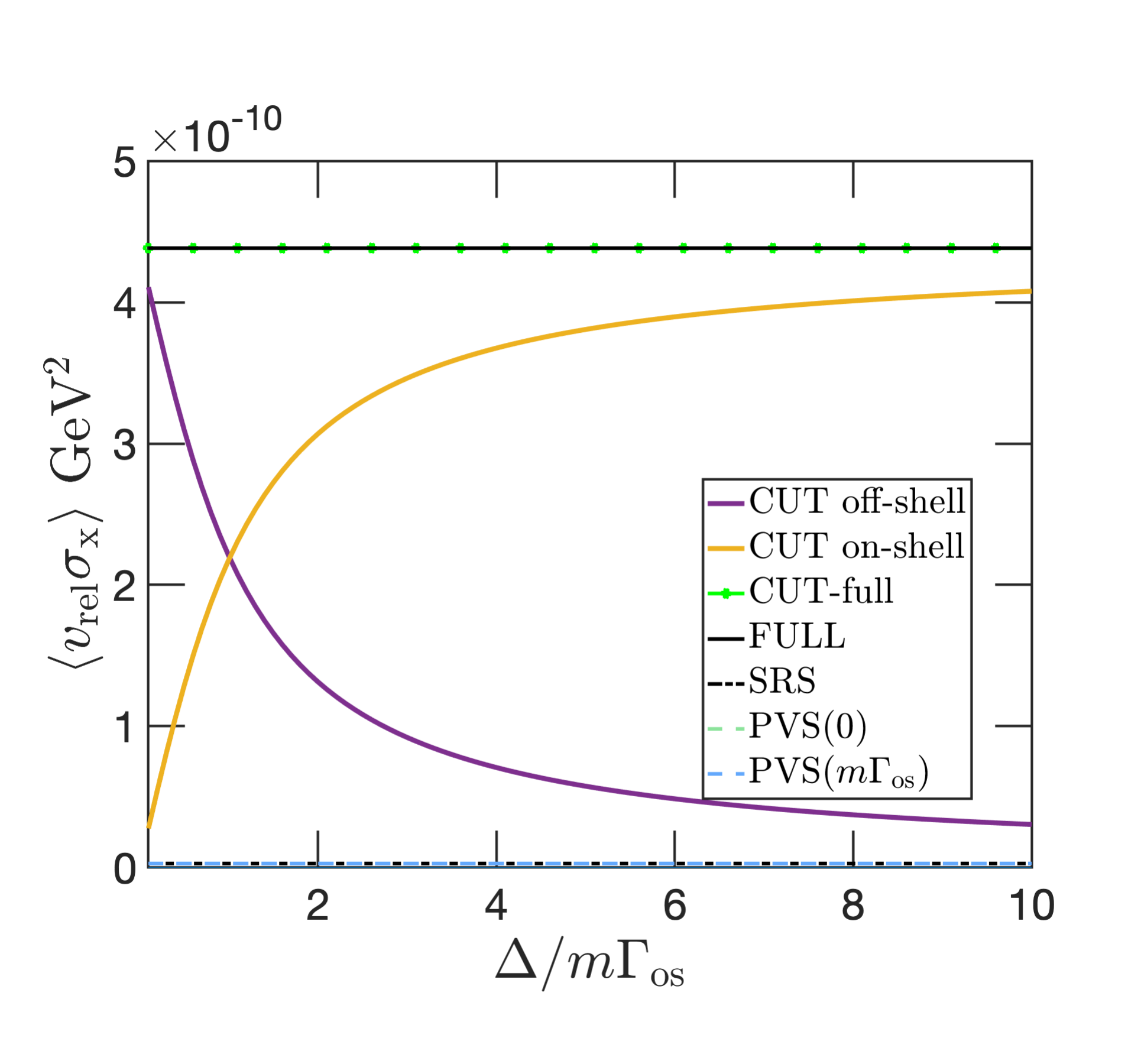}
\vskip-0.3cm
\caption{The solid purple curve corresponds to the off-shell averaged cross section in the cut scheme, Eq.~\cref{eq:off-shell_cut}, and the solid yellow curve shows to the on-shell result, Eq.~\eqref{eq:on-shell_cut}. The black line is the full integral computed with the BW propagator. We used the same parameters as in Fig.~\cref{fig:offshell_comparison}.} 
\label{fig:cut_comparison_1}
\end{figure}
%

In Fig.~\cref{fig:cut_comparison_1} we show the thermal averaged results in the cut scheme, Eqs.~\cref{eq:cut_scheme} and~\cref{eq:on-shell_cut}, as a function of the width of the cut area $\Delta$. The solid purple line presents the off-shell scattering contribution, and the solid yellow line shows the on-shell part. The solid black line shows the full thermal integral, which here agrees with the sum of the off- and on-shell cut contributions. Note that a relatively large $\Delta$ is needed to make the off-shell cut result agree with the standard subtraction schemes, or equivalently to have $R(\Delta) \approx 1$. Indeed, to get $R(\Delta) = 0.99$, one needs $\Delta = 70 \epsilon_{\rm os}$. But even then one finds $1-r(70 \epsilon_{\rm os}) \approx 4\times 10^{-4}$, where $r(\Delta)$ is the sum of the two cut contributions divided by the full result. This shows that the effective particle approximation works very well. This is as expected since the resonance is very narrow $\Gamma_{\rm os}/T \approx 0.01$. The difference between the standard subtraction schemes is also not visible in this scale.

%
\begin{figure}
\includegraphics[width=0.38\textwidth]{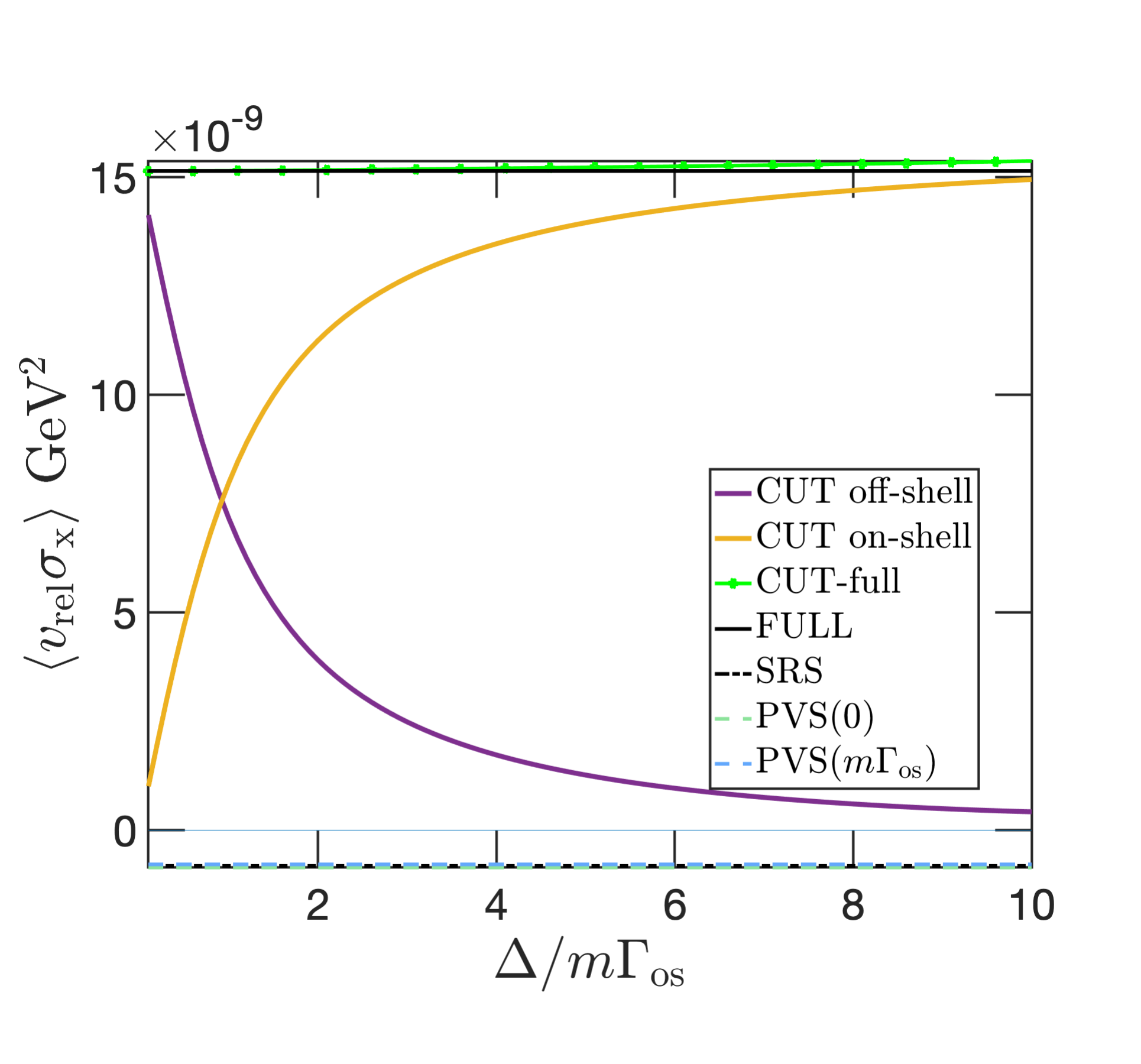}
\vskip-0.2cm
\caption{Same as in Fig.~\cref{fig:cut_comparison_1}, but for parameters $\lambda_{hs}= 0.01$ and $T = 50$ GeV. Note that the standard subtraction scheme results are all negative in this case.} 
\label{fig:cut_comparison_2}
\end{figure}
%

\begin{figure}
\includegraphics[width=0.38\textwidth]{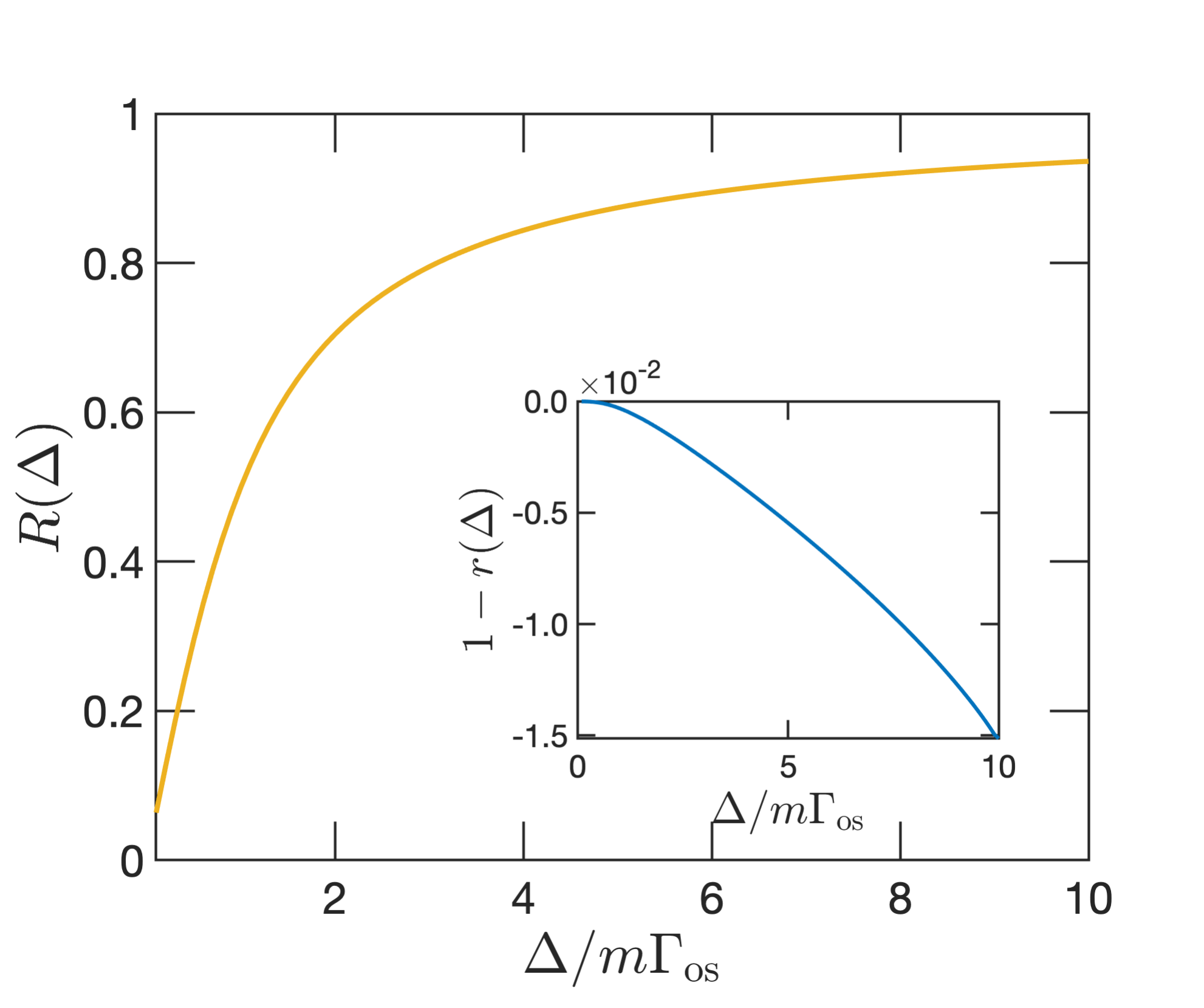}
\vskip-0.2cm
\caption{Shown is the weight function $R(\Delta)$ for the parameters used in Fig.~\cref{fig:cut_comparison_2}. The inset shows the deviation of the sum of cut contributions from the full thermal integral, as explained in the text.} 
\label{fig:weight_2}
\end{figure}

In Fig.~\cref{fig:cut_comparison_2} we show results of a computation with a larger coupling $\lambda_{hs} = 0.01$ and at a higher temperature $T = 50$ GeV. The variation between different PVS and SRS schemes is about 10\% here, which is still too small to be visible in the plot, but we see that all schemes give {\em negative} scattering cross sections. However, the sum of the on-shell and off-shell contributions still is very close to (and in the SRS case exactly) the full result. This demonstrates what we already stated above: both SRS and PVS schemes overestimate the on-shell parts, which then must be compensated by unphysical negative scattering contributions. In the cut case all contributions remain positive of course. However, we now have a larger width $\Gamma_{\rm os}/T \approx 0.09$, which implies that one cannot choose as large $\Delta$ as in the first example. The deviation of the cut contributions from the exact result, given by $1-r(\Delta)$, is shown in the inset of Fig.~\cref{fig:weight_2}. We see that already for $\Delta = 10\epsilon_{\rm os}$ we have a percent level deviation from the accurate result, while $R(10\epsilon_{\rm os}) \approx 0.93$ is still well below unity. In this case the best spectral modeling of the system would correspond to a cut scheme with $\Delta \approx 5$ and $R \approx 0.9$; \eg~there should be a suppression in the decay channel contributions to the collision integrals.

\begin{figure}
\includegraphics[width=0.38\textwidth]{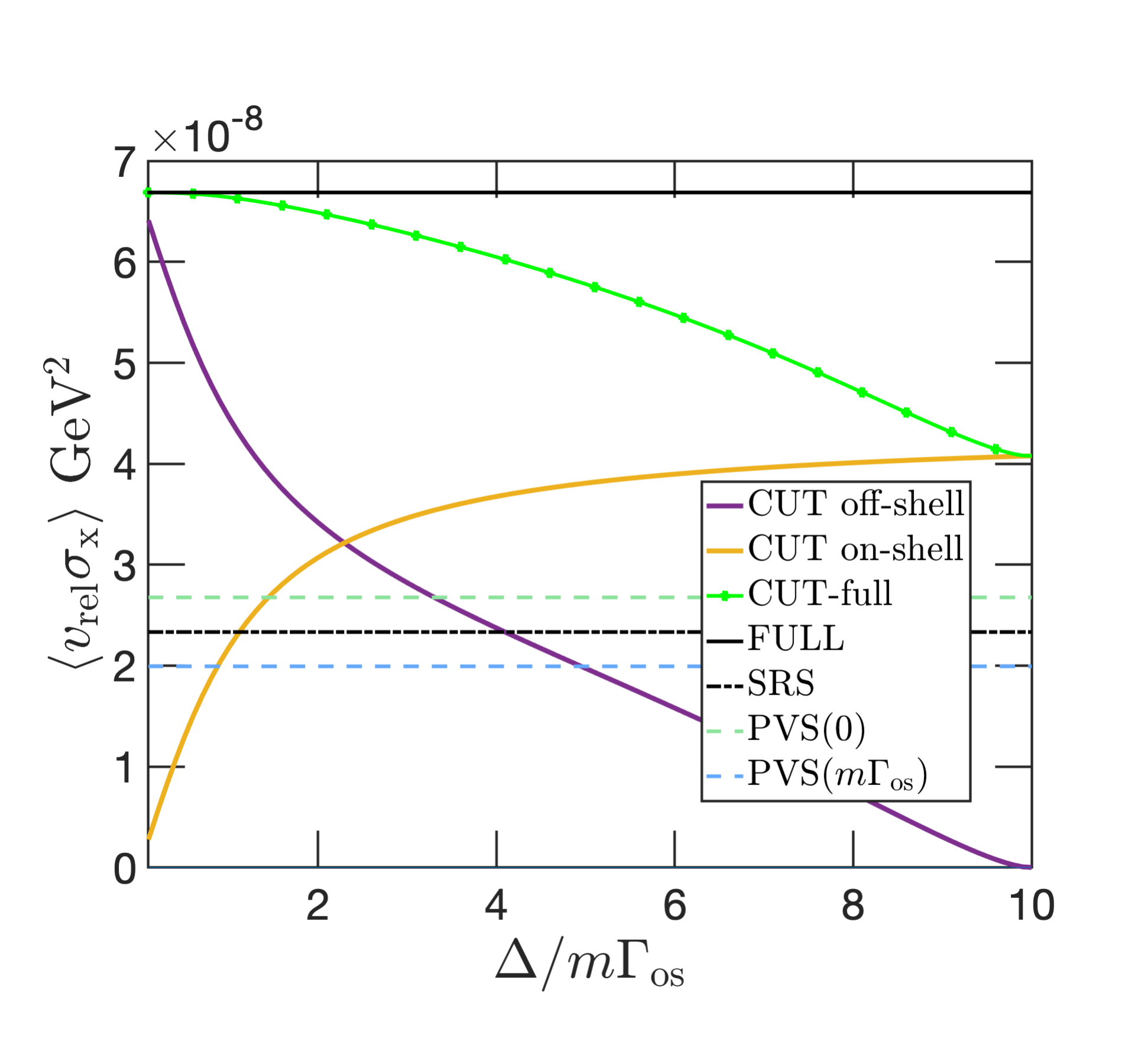}
\vskip-0.2cm
\caption{The same as in Fig.~\cref{fig:cut_comparison_2}, but now in the temperature
$T = 5$ GeV.} 
\label{fig:cut_comparison_3}
\end{figure}

Finally, in Fig.~\cref{fig:cut_comparison_3} we present results for the same case as in Fig.~\cref{fig:cut_comparison_2}, but at a lower temperature of $T = 5$ GeV. Here the SRS and the PVS scenarios again give positive results with a dispersion of about 25\%. One might then think that the division to on-shell and off-shell contributions still makes sense, but this is not the case, as can be seen from the cut-scheme results. The sum of the cut contributions starts to deviate from the full result already when $\Delta > \epsilon_{\rm os}$ and the off-shell contribution gets completely erased before the on-shell result reaches its maximum. In this case the Higgs resonance cannot be treated as an on-shell particle. Or conversely, if one insists in doing so, one should use $\Delta < \epsilon_{\rm os}$, giving strong suppression to the decay contributions: $R < 0.5$.

%
%

%
%

To conclude this section we note that none of the scenarios discussed above was realistic in the sense that one {\em needed} to deal with the problem of on- and off-shell division. In the singlet model the accurate evaluation of the final dark matter abundance only requires that the kinetic description is valid near the freeze-out temperature, which does not require a kinetic equation for the on-shell excitations of the Higgs field. There are other realistic settings~\cite{Cline:2017qpe,Merle:2015oja,Ala-Mattinen:2022nuj}, where these effects may be relevant, however, and our examples display qualitatively the problems one might encounter. Overall the standard schemes work rather well and are indistinguishable when applied correctly. Moreover, even the negative cross sections are not necessarily fatal, although the accuracy of schemes leading to them is difficult to assess without more detailed computations.

%
\section{Conclusions}
\label{sec:conclusions}
%

In this article we revisited the well-known RIS subtraction problem. The RIS subtraction is necessary for the resonant scattering channels when the intermediate state causing the resonance is also included as a real state in the on-shell kinetic equation network. Over the years many different methods to perform the RIS subtraction have been proposed, with some of them leading to erroneous results and sometimes to negative scattering cross sections. 

We took a fresh look into the different RIS schemes that have appeared in the literature. We pointed out how to extend the standard RIS-subtraction scheme to interference matrix elements and emphasized the need to use the correct sequences for different propagator functions in the principal value scheme. It indeed seems that most errors in the literature associated with these schemes have arisen from not realizing that the off-shell propagator and the squared off-shell propagator need separate definitions.

We then discussed the origin of RIS problem from the Schwinger-Dyson equation point of view. The freedom in the definition of the RIS-subtracted propagators corresponds to a freedom in defining the Hermitian part of the pole functions when Schwinger-Dyson equations for unstable particles are reduced to the spectral limit. We used this insight to propose a more physical cut-subtraction scheme, where the modes within distance $\Delta$ from the resonance are associated with the on-shell state and those outside this interval are treated as virtual states. This method avoids the negative cross sections and provides a direct measure, in terms of an on-shell weight function $\w(\Delta)$, to estimate when the on-shell particle picture is a consistent description. 

We finished by a detailed numerical comparison of the different subtraction schemes in a toy model which qualitatively reproduces the various issues that can be encountered with the different subtraction schemes. We pointed out that the negative cross sections in the SRS and PVS schemes, while unphysical at face value, arise to compensate for the overestimated on-shell contributions in these schemes. This problem never appears in the cut scheme. It would obviously be interesting to make detailed comparisons of the different approaches in more realistic setups and with full out-of-equilibrium computations.

\vskip-1.7cm

%
\section*{ACKNOWLEDGEMENTS}
This work was supported by the Research Council of Finland Grant No.~342777 and Jenny and Antti Wihuri Foundation.
%

%
\bibliography{RIS}

\begin{thebibliography}{44}
\expandafter\ifx\csname natexlab\endcsname\relax\def\natexlab#1{#1}\fi
\expandafter\ifx\csname bibnamefont\endcsname\relax
  \def\bibnamefont#1{#1}\fi
\expandafter\ifx\csname bibfnamefont\endcsname\relax
  \def\bibfnamefont#1{#1}\fi
\expandafter\ifx\csname citenamefont\endcsname\relax
  \def\citenamefont#1{#1}\fi
\expandafter\ifx\csname url\endcsname\relax
  \def\url#1{\texttt{#1}}\fi
\expandafter\ifx\csname urlprefix\endcsname\relax\def\urlprefix{URL }\fi
\providecommand{\bibinfo}[2]{#2}
\providecommand{\eprint}[2][]{\url{#2}}

\bibitem[{\citenamefont{Kolb and Wolfram}(1980{\natexlab{a}})}]{Kolb:1979ui}
\bibinfo{author}{\bibfnamefont{E.~W.} \bibnamefont{Kolb}} \bibnamefont{and} \bibinfo{author}{\bibfnamefont{S.}~\bibnamefont{Wolfram}}, \bibinfo{journal}{Phys. Lett. B} \textbf{\bibinfo{volume}{91}}, \bibinfo{pages}{217} (\bibinfo{year}{1980}{\natexlab{a}}).

\bibitem[{\citenamefont{Kolb and Wolfram}(1980{\natexlab{b}})}]{Kolb:1979qa}
\bibinfo{author}{\bibfnamefont{E.~W.} \bibnamefont{Kolb}} \bibnamefont{and} \bibinfo{author}{\bibfnamefont{S.}~\bibnamefont{Wolfram}}, \bibinfo{journal}{Nucl. Phys. B} \textbf{\bibinfo{volume}{172}}, \bibinfo{pages}{224} (\bibinfo{year}{1980}{\natexlab{b}}), \bibinfo{note}{[Erratum: Nucl.Phys.B 195, 542 (1982)]}.

\bibitem[{\citenamefont{Fry et~al.}(1980)\citenamefont{Fry, Olive, and Turner}}]{Fry:1980ph}
\bibinfo{author}{\bibfnamefont{J.~N.} \bibnamefont{Fry}}, \bibinfo{author}{\bibfnamefont{K.~A.} \bibnamefont{Olive}}, \bibnamefont{and} \bibinfo{author}{\bibfnamefont{M.~S.} \bibnamefont{Turner}}, \bibinfo{journal}{Phys. Rev. D} \textbf{\bibinfo{volume}{22}}, \bibinfo{pages}{2953} (\bibinfo{year}{1980}).

\bibitem[{\citenamefont{Harvey et~al.}(1982)\citenamefont{Harvey, Kolb, Reiss, and Wolfram}}]{Harvey:1981yk}
\bibinfo{author}{\bibfnamefont{J.~A.} \bibnamefont{Harvey}}, \bibinfo{author}{\bibfnamefont{E.~W.} \bibnamefont{Kolb}}, \bibinfo{author}{\bibfnamefont{D.~B.} \bibnamefont{Reiss}}, \bibnamefont{and} \bibinfo{author}{\bibfnamefont{S.}~\bibnamefont{Wolfram}}, \bibinfo{journal}{Nucl. Phys. B} \textbf{\bibinfo{volume}{201}}, \bibinfo{pages}{16} (\bibinfo{year}{1982}).

\bibitem[{\citenamefont{Luty}(1992)}]{Luty:1992un}
\bibinfo{author}{\bibfnamefont{M.~A.} \bibnamefont{Luty}}, \bibinfo{journal}{Phys. Rev. D} \textbf{\bibinfo{volume}{45}}, \bibinfo{pages}{455} (\bibinfo{year}{1992}).

\bibitem[{\citenamefont{Cline et~al.}(1994)\citenamefont{Cline, Kainulainen, and Olive}}]{Cline:1993bd}
\bibinfo{author}{\bibfnamefont{J.~M.} \bibnamefont{Cline}}, \bibinfo{author}{\bibfnamefont{K.}~\bibnamefont{Kainulainen}}, \bibnamefont{and} \bibinfo{author}{\bibfnamefont{K.~A.} \bibnamefont{Olive}}, \bibinfo{journal}{Phys. Rev. D} \textbf{\bibinfo{volume}{49}}, \bibinfo{pages}{6394} (\bibinfo{year}{1994}), \eprint{hep-ph/9401208}.

\bibitem[{\citenamefont{Bouzas et~al.}(1997)\citenamefont{Bouzas, Cheng, and Gelmini}}]{Bouzas:1996jg}
\bibinfo{author}{\bibfnamefont{A.~O.} \bibnamefont{Bouzas}}, \bibinfo{author}{\bibfnamefont{W.~Y.} \bibnamefont{Cheng}}, \bibnamefont{and} \bibinfo{author}{\bibfnamefont{G.}~\bibnamefont{Gelmini}}, \bibinfo{journal}{Phys. Rev. D} \textbf{\bibinfo{volume}{55}}, \bibinfo{pages}{4663} (\bibinfo{year}{1997}), \eprint{hep-ph/9602278}.

\bibitem[{\citenamefont{Plumacher}(1997)}]{Plumacher:1996kc}
\bibinfo{author}{\bibfnamefont{M.}~\bibnamefont{Plumacher}}, \bibinfo{journal}{Z. Phys. C} \textbf{\bibinfo{volume}{74}}, \bibinfo{pages}{549} (\bibinfo{year}{1997}), \eprint{hep-ph/9604229}.

\bibitem[{\citenamefont{Papavassiliou and Pilaftsis}(1996)}]{Papavassiliou:1996zn}
\bibinfo{author}{\bibfnamefont{J.}~\bibnamefont{Papavassiliou}} \bibnamefont{and} \bibinfo{author}{\bibfnamefont{A.}~\bibnamefont{Pilaftsis}}, \bibinfo{journal}{Phys. Rev. D} \textbf{\bibinfo{volume}{54}}, \bibinfo{pages}{5315} (\bibinfo{year}{1996}), \eprint{hep-ph/9605385}.

\bibitem[{\citenamefont{Pilaftsis}(1997)}]{Pilaftsis:1997jf}
\bibinfo{author}{\bibfnamefont{A.}~\bibnamefont{Pilaftsis}}, \bibinfo{journal}{Phys. Rev. D} \textbf{\bibinfo{volume}{56}}, \bibinfo{pages}{5431} (\bibinfo{year}{1997}), \eprint{hep-ph/9707235}.

\bibitem[{\citenamefont{Hambye et~al.}(2000)\citenamefont{Hambye, Ma, and Sarkar}}]{Hambye:2000zs}
\bibinfo{author}{\bibfnamefont{T.}~\bibnamefont{Hambye}}, \bibinfo{author}{\bibfnamefont{E.}~\bibnamefont{Ma}}, \bibnamefont{and} \bibinfo{author}{\bibfnamefont{U.}~\bibnamefont{Sarkar}}, \bibinfo{journal}{Nucl. Phys. B} \textbf{\bibinfo{volume}{590}}, \bibinfo{pages}{429} (\bibinfo{year}{2000}), \eprint{hep-ph/0006173}.

\bibitem[{\citenamefont{Giudice et~al.}(2004)\citenamefont{Giudice, Notari, Raidal, Riotto, and Strumia}}]{Giudice:2003jh}
\bibinfo{author}{\bibfnamefont{G.~F.} \bibnamefont{Giudice}}, \bibinfo{author}{\bibfnamefont{A.}~\bibnamefont{Notari}}, \bibinfo{author}{\bibfnamefont{M.}~\bibnamefont{Raidal}}, \bibinfo{author}{\bibfnamefont{A.}~\bibnamefont{Riotto}}, \bibnamefont{and} \bibinfo{author}{\bibfnamefont{A.}~\bibnamefont{Strumia}}, \bibinfo{journal}{Nucl. Phys. B} \textbf{\bibinfo{volume}{685}}, \bibinfo{pages}{89} (\bibinfo{year}{2004}), \eprint{hep-ph/0310123}.

\bibitem[{\citenamefont{Pilaftsis and Underwood}(2004)}]{Pilaftsis:2003gt}
\bibinfo{author}{\bibfnamefont{A.}~\bibnamefont{Pilaftsis}} \bibnamefont{and} \bibinfo{author}{\bibfnamefont{T.~E.~J.} \bibnamefont{Underwood}}, \bibinfo{journal}{Nucl. Phys. B} \textbf{\bibinfo{volume}{692}}, \bibinfo{pages}{303} (\bibinfo{year}{2004}), \eprint{hep-ph/0309342}.

\bibitem[{\citenamefont{Sawyer}(2004)}]{Sawyer:2003yi}
\bibinfo{author}{\bibfnamefont{R.~F.} \bibnamefont{Sawyer}}, \bibinfo{journal}{Phys. Rev. D} \textbf{\bibinfo{volume}{69}}, \bibinfo{pages}{083001} (\bibinfo{year}{2004}), \eprint{hep-ph/0312158}.

\bibitem[{\citenamefont{Buchmuller et~al.}(2005)\citenamefont{Buchmuller, Di~Bari, and Plumacher}}]{Buchmuller:2004nz}
\bibinfo{author}{\bibfnamefont{W.}~\bibnamefont{Buchmuller}}, \bibinfo{author}{\bibfnamefont{P.}~\bibnamefont{Di~Bari}}, \bibnamefont{and} \bibinfo{author}{\bibfnamefont{M.}~\bibnamefont{Plumacher}}, \bibinfo{journal}{Annals Phys.} \textbf{\bibinfo{volume}{315}}, \bibinfo{pages}{305} (\bibinfo{year}{2005}), \eprint{hep-ph/0401240}.

\bibitem[{\citenamefont{Nardi et~al.}(2007)\citenamefont{Nardi, Racker, and Roulet}}]{Nardi:2007jp}
\bibinfo{author}{\bibfnamefont{E.}~\bibnamefont{Nardi}}, \bibinfo{author}{\bibfnamefont{J.}~\bibnamefont{Racker}}, \bibnamefont{and} \bibinfo{author}{\bibfnamefont{E.}~\bibnamefont{Roulet}}, \bibinfo{journal}{JHEP} \textbf{\bibinfo{volume}{09}}, \bibinfo{pages}{090} (\bibinfo{year}{2007}), \eprint{0707.0378}.

\bibitem[{\citenamefont{Davidson et~al.}(2008)\citenamefont{Davidson, Nardi, and Nir}}]{Davidson:2008bu}
\bibinfo{author}{\bibfnamefont{S.}~\bibnamefont{Davidson}}, \bibinfo{author}{\bibfnamefont{E.}~\bibnamefont{Nardi}}, \bibnamefont{and} \bibinfo{author}{\bibfnamefont{Y.}~\bibnamefont{Nir}}, \bibinfo{journal}{Phys. Rept.} \textbf{\bibinfo{volume}{466}}, \bibinfo{pages}{105} (\bibinfo{year}{2008}), \eprint{0802.2962}.

\bibitem[{\citenamefont{Deppisch and Pilaftsis}(2011)}]{Deppisch:2010fr}
\bibinfo{author}{\bibfnamefont{F.~F.} \bibnamefont{Deppisch}} \bibnamefont{and} \bibinfo{author}{\bibfnamefont{A.}~\bibnamefont{Pilaftsis}}, \bibinfo{journal}{Phys. Rev. D} \textbf{\bibinfo{volume}{83}}, \bibinfo{pages}{076007} (\bibinfo{year}{2011}), \eprint{1012.1834}.

\bibitem[{\citenamefont{Kiessig and Plumacher}(2012)}]{Kiessig:2011ga}
\bibinfo{author}{\bibfnamefont{C.}~\bibnamefont{Kiessig}} \bibnamefont{and} \bibinfo{author}{\bibfnamefont{M.}~\bibnamefont{Plumacher}}, \bibinfo{journal}{JCAP} \textbf{\bibinfo{volume}{09}}, \bibinfo{pages}{012} (\bibinfo{year}{2012}), \eprint{1111.1235}.

\bibitem[{\citenamefont{Frossard et~al.}(2013)\citenamefont{Frossard, Garny, Hohenegger, Kartavtsev, and Mitrouskas}}]{Frossard:2012pc}
\bibinfo{author}{\bibfnamefont{T.}~\bibnamefont{Frossard}}, \bibinfo{author}{\bibfnamefont{M.}~\bibnamefont{Garny}}, \bibinfo{author}{\bibfnamefont{A.}~\bibnamefont{Hohenegger}}, \bibinfo{author}{\bibfnamefont{A.}~\bibnamefont{Kartavtsev}}, \bibnamefont{and} \bibinfo{author}{\bibfnamefont{D.}~\bibnamefont{Mitrouskas}}, \bibinfo{journal}{Phys. Rev. D} \textbf{\bibinfo{volume}{87}}, \bibinfo{pages}{085009} (\bibinfo{year}{2013}), \eprint{1211.2140}.

\bibitem[{\citenamefont{Iso et~al.}(2014)\citenamefont{Iso, Shimada, and Yamanaka}}]{Iso:2013lba}
\bibinfo{author}{\bibfnamefont{S.}~\bibnamefont{Iso}}, \bibinfo{author}{\bibfnamefont{K.}~\bibnamefont{Shimada}}, \bibnamefont{and} \bibinfo{author}{\bibfnamefont{M.}~\bibnamefont{Yamanaka}}, \bibinfo{journal}{JHEP} \textbf{\bibinfo{volume}{04}}, \bibinfo{pages}{062} (\bibinfo{year}{2014}), \eprint{1312.7680}.

\bibitem[{\citenamefont{Iso and Shimada}(2014)}]{Iso:2014afa}
\bibinfo{author}{\bibfnamefont{S.}~\bibnamefont{Iso}} \bibnamefont{and} \bibinfo{author}{\bibfnamefont{K.}~\bibnamefont{Shimada}}, \bibinfo{journal}{JHEP} \textbf{\bibinfo{volume}{08}}, \bibinfo{pages}{043} (\bibinfo{year}{2014}), \eprint{1404.4816}.

\bibitem[{\citenamefont{Lavignac and Schmauch}(2015)}]{Lavignac:2015gpa}
\bibinfo{author}{\bibfnamefont{S.}~\bibnamefont{Lavignac}} \bibnamefont{and} \bibinfo{author}{\bibfnamefont{B.}~\bibnamefont{Schmauch}}, \bibinfo{journal}{JHEP} \textbf{\bibinfo{volume}{05}}, \bibinfo{pages}{124} (\bibinfo{year}{2015}), \eprint{1503.00629}.

\bibitem[{\citenamefont{Cline et~al.}(2017)\citenamefont{Cline, Kainulainen, and Tucker-Smith}}]{Cline:2017qpe}
\bibinfo{author}{\bibfnamefont{J.~M.} \bibnamefont{Cline}}, \bibinfo{author}{\bibfnamefont{K.}~\bibnamefont{Kainulainen}}, \bibnamefont{and} \bibinfo{author}{\bibfnamefont{D.}~\bibnamefont{Tucker-Smith}}, \bibinfo{journal}{Phys. Rev. D} \textbf{\bibinfo{volume}{95}}, \bibinfo{pages}{115006} (\bibinfo{year}{2017}), \eprint{1702.08909}.

\bibitem[{\citenamefont{Biondini et~al.}(2018)}]{Biondini:2017rpb}
\bibinfo{author}{\bibfnamefont{S.}~\bibnamefont{Biondini}} \bibnamefont{et~al.}, \bibinfo{journal}{Int. J. Mod. Phys. A} \textbf{\bibinfo{volume}{33}}, \bibinfo{pages}{1842004} (\bibinfo{year}{2018}), \eprint{1711.02864}.

\bibitem[{\citenamefont{Dev et~al.}(2018)\citenamefont{Dev, Di~Bari, Garbrecht, Lavignac, Millington, and Teresi}}]{Dev:2017trv}
\bibinfo{author}{\bibfnamefont{P.~S.~B.} \bibnamefont{Dev}}, \bibinfo{author}{\bibfnamefont{P.}~\bibnamefont{Di~Bari}}, \bibinfo{author}{\bibfnamefont{B.}~\bibnamefont{Garbrecht}}, \bibinfo{author}{\bibfnamefont{S.}~\bibnamefont{Lavignac}}, \bibinfo{author}{\bibfnamefont{P.}~\bibnamefont{Millington}}, \bibnamefont{and} \bibinfo{author}{\bibfnamefont{D.}~\bibnamefont{Teresi}}, \bibinfo{journal}{Int. J. Mod. Phys. A} \textbf{\bibinfo{volume}{33}}, \bibinfo{pages}{1842001} (\bibinfo{year}{2018}), \eprint{1711.02861}.

\bibitem[{\citenamefont{Bodeker and Buchmuller}(2021)}]{Bodeker:2020ghk}
\bibinfo{author}{\bibfnamefont{D.}~\bibnamefont{Bodeker}} \bibnamefont{and} \bibinfo{author}{\bibfnamefont{W.}~\bibnamefont{Buchmuller}}, \bibinfo{journal}{Rev. Mod. Phys.} \textbf{\bibinfo{volume}{93}}, \bibinfo{pages}{035004} (\bibinfo{year}{2021}), \eprint{2009.07294}.

\bibitem[{\citenamefont{Matak}(2022)}]{Matak:2022qwc}
\bibinfo{author}{\bibfnamefont{P.}~\bibnamefont{Matak}}, \bibinfo{journal}{Phys. Rev. D} \textbf{\bibinfo{volume}{105}}, \bibinfo{pages}{076019} (\bibinfo{year}{2022}), \eprint{2203.01253}.

\bibitem[{\citenamefont{Mat\'ak}(2023)}]{Matak:2023zox}
\bibinfo{author}{\bibfnamefont{P.}~\bibnamefont{Mat\'ak}} (\bibinfo{year}{2023}), \eprint{2305.19238}.

\bibitem[{\citenamefont{Ala-Mattinen et~al.}(2022)\citenamefont{Ala-Mattinen, Heikinheimo, Kainulainen, and Tuominen}}]{Ala-Mattinen:2022nuj}
\bibinfo{author}{\bibfnamefont{K.}~\bibnamefont{Ala-Mattinen}}, \bibinfo{author}{\bibfnamefont{M.}~\bibnamefont{Heikinheimo}}, \bibinfo{author}{\bibfnamefont{K.}~\bibnamefont{Kainulainen}}, \bibnamefont{and} \bibinfo{author}{\bibfnamefont{K.}~\bibnamefont{Tuominen}}, \bibinfo{journal}{Phys. Rev. D} \textbf{\bibinfo{volume}{105}}, \bibinfo{pages}{123005} (\bibinfo{year}{2022}), \eprint{2201.06456}.

\bibitem[{\citenamefont{Laine}(2023)}]{Laine:2022ner}
\bibinfo{author}{\bibfnamefont{M.}~\bibnamefont{Laine}}, \bibinfo{journal}{JHEP} \textbf{\bibinfo{volume}{01}}, \bibinfo{pages}{157} (\bibinfo{year}{2023}), \eprint{2211.06008}.

\bibitem[{\citenamefont{Jukkala et~al.}(2021)\citenamefont{Jukkala, Kainulainen, and Rahkila}}]{Juk21}
\bibinfo{author}{\bibfnamefont{H.}~\bibnamefont{Jukkala}}, \bibinfo{author}{\bibfnamefont{K.}~\bibnamefont{Kainulainen}}, \bibnamefont{and} \bibinfo{author}{\bibfnamefont{P.~M.} \bibnamefont{Rahkila}}, \bibinfo{journal}{JHEP} \textbf{\bibinfo{volume}{09}}, \bibinfo{pages}{119} (\bibinfo{year}{2021}), \eprint{2104.03998}.

\bibitem[{\citenamefont{Kainulainen and Parkkinen}(2023{\natexlab{a}})}]{Kainulainen:2023ocv}
\bibinfo{author}{\bibfnamefont{K.}~\bibnamefont{Kainulainen}} \bibnamefont{and} \bibinfo{author}{\bibfnamefont{H.}~\bibnamefont{Parkkinen}} (\bibinfo{year}{2023}{\natexlab{a}}), \eprint{2309.00881}.

\bibitem[{\citenamefont{Kainulainen and Parkkinen}(2023{\natexlab{b}})}]{Kainulainen:2023khg}
\bibinfo{author}{\bibfnamefont{K.}~\bibnamefont{Kainulainen}} \bibnamefont{and} \bibinfo{author}{\bibfnamefont{H.~H.} \bibnamefont{Parkkinen}}, \bibinfo{journal}{PoS} \textbf{\bibinfo{volume}{ICRC2023}}, \bibinfo{pages}{1127} (\bibinfo{year}{2023}{\natexlab{b}}).

\bibitem[{\citenamefont{Silveira and Zee}(1985)}]{Silveira:1985rk}
\bibinfo{author}{\bibfnamefont{V.}~\bibnamefont{Silveira}} \bibnamefont{and} \bibinfo{author}{\bibfnamefont{A.}~\bibnamefont{Zee}}, \bibinfo{journal}{Phys. Lett. B} \textbf{\bibinfo{volume}{161}}, \bibinfo{pages}{136} (\bibinfo{year}{1985}).

\bibitem[{\citenamefont{McDonald}(1994)}]{McDonald:1993ex}
\bibinfo{author}{\bibfnamefont{J.}~\bibnamefont{McDonald}}, \bibinfo{journal}{Phys. Rev. D} \textbf{\bibinfo{volume}{50}}, \bibinfo{pages}{3637} (\bibinfo{year}{1994}), \eprint{hep-ph/0702143}.

\bibitem[{\citenamefont{Burgess et~al.}(2001)\citenamefont{Burgess, Pospelov, and ter Veldhuis}}]{Burgess:2000yq}
\bibinfo{author}{\bibfnamefont{C.~P.} \bibnamefont{Burgess}}, \bibinfo{author}{\bibfnamefont{M.}~\bibnamefont{Pospelov}}, \bibnamefont{and} \bibinfo{author}{\bibfnamefont{T.}~\bibnamefont{ter Veldhuis}}, \bibinfo{journal}{Nucl. Phys. B} \textbf{\bibinfo{volume}{619}}, \bibinfo{pages}{709} (\bibinfo{year}{2001}), \eprint{hep-ph/0011335}.

\bibitem[{\citenamefont{Barger et~al.}(2008)\citenamefont{Barger, Langacker, McCaskey, Ramsey-Musolf, and Shaughnessy}}]{Barger:2007im}
\bibinfo{author}{\bibfnamefont{V.}~\bibnamefont{Barger}}, \bibinfo{author}{\bibfnamefont{P.}~\bibnamefont{Langacker}}, \bibinfo{author}{\bibfnamefont{M.}~\bibnamefont{McCaskey}}, \bibinfo{author}{\bibfnamefont{M.~J.} \bibnamefont{Ramsey-Musolf}}, \bibnamefont{and} \bibinfo{author}{\bibfnamefont{G.}~\bibnamefont{Shaughnessy}}, \bibinfo{journal}{Phys. Rev. D} \textbf{\bibinfo{volume}{77}}, \bibinfo{pages}{035005} (\bibinfo{year}{2008}), \eprint{0706.4311}.

\bibitem[{\citenamefont{Cline et~al.}(2013)\citenamefont{Cline, Kainulainen, Scott, and Weniger}}]{Cline:2013gha}
\bibinfo{author}{\bibfnamefont{J.~M.} \bibnamefont{Cline}}, \bibinfo{author}{\bibfnamefont{K.}~\bibnamefont{Kainulainen}}, \bibinfo{author}{\bibfnamefont{P.}~\bibnamefont{Scott}}, \bibnamefont{and} \bibinfo{author}{\bibfnamefont{C.}~\bibnamefont{Weniger}}, \bibinfo{journal}{Phys. Rev. D} \textbf{\bibinfo{volume}{88}}, \bibinfo{pages}{055025} (\bibinfo{year}{2013}), \bibinfo{note}{[Erratum: Phys.Rev.D 92, 039906 (2015)]}, \eprint{1306.4710}.

\bibitem[{\citenamefont{Binder et~al.}(2017)\citenamefont{Binder, Bringmann, Gustafsson, and Hryczuk}}]{Binder:2017rgn}
\bibinfo{author}{\bibfnamefont{T.}~\bibnamefont{Binder}}, \bibinfo{author}{\bibfnamefont{T.}~\bibnamefont{Bringmann}}, \bibinfo{author}{\bibfnamefont{M.}~\bibnamefont{Gustafsson}}, \bibnamefont{and} \bibinfo{author}{\bibfnamefont{A.}~\bibnamefont{Hryczuk}}, \bibinfo{journal}{Phys. Rev. D} \textbf{\bibinfo{volume}{96}}, \bibinfo{pages}{115010} (\bibinfo{year}{2017}), \bibinfo{note}{[Erratum: Phys.Rev.D 101, 099901 (2020)]}, \eprint{1706.07433}.

\bibitem[{\citenamefont{Ala-Mattinen and Kainulainen}(2020)}]{Ala-Mattinen:2019mpa}
\bibinfo{author}{\bibfnamefont{K.}~\bibnamefont{Ala-Mattinen}} \bibnamefont{and} \bibinfo{author}{\bibfnamefont{K.}~\bibnamefont{Kainulainen}}, \bibinfo{journal}{JCAP} \textbf{\bibinfo{volume}{09}}, \bibinfo{pages}{040} (\bibinfo{year}{2020}), \eprint{1912.02870}.

\bibitem[{\citenamefont{Kim and Laine}(2023)}]{Kim:2023bxy}
\bibinfo{author}{\bibfnamefont{S.}~\bibnamefont{Kim}} \bibnamefont{and} \bibinfo{author}{\bibfnamefont{M.}~\bibnamefont{Laine}}, \bibinfo{journal}{JCAP} \textbf{\bibinfo{volume}{05}}, \bibinfo{pages}{003} (\bibinfo{year}{2023}), \eprint{2302.05129}.

\bibitem[{\citenamefont{Gondolo and Gelmini}(1991)}]{Gondolo:1990dk}
\bibinfo{author}{\bibfnamefont{P.}~\bibnamefont{Gondolo}} \bibnamefont{and} \bibinfo{author}{\bibfnamefont{G.}~\bibnamefont{Gelmini}}, \bibinfo{journal}{Nucl. Phys. B} \textbf{\bibinfo{volume}{360}}, \bibinfo{pages}{145} (\bibinfo{year}{1991}).

\bibitem[{\citenamefont{Merle and Totzauer}(2015)}]{Merle:2015oja}
\bibinfo{author}{\bibfnamefont{A.}~\bibnamefont{Merle}} \bibnamefont{and} \bibinfo{author}{\bibfnamefont{M.}~\bibnamefont{Totzauer}}, \bibinfo{journal}{JCAP} \textbf{\bibinfo{volume}{1506}}, \bibinfo{pages}{011} (\bibinfo{year}{2015}), \eprint{1502.01011}.

\end{thebibliography}
%

%
\end{document}